\documentclass[10pt,twocolumn,letterpaper,shortlabels]{article}

\usepackage[pagenumbers]{cvpr} 

\pdfoutput=1
\usepackage{lineno}
\usepackage{lipsum}

\usepackage[whole]{bxcjkjatype}
\usepackage{color}
\usepackage{amsmath}
\usepackage{amssymb}
\usepackage{amsfonts}
\usepackage{pifont} 
\usepackage{bbding} 
\usepackage[rgb,dvipsnames]{xcolor}
\usepackage{makecell}
\newcommand{\norm}[1]{\left\lVert#1\right\rVert}
\usepackage{mathrsfs}
\usepackage{array}

\newcommand{\redcross}{\textcolor{red}{\XSolidBrush}}
\newcommand{\greencheck}{\textcolor{green}{\checkmark}}

\definecolor{cvprblue}{rgb}{0.21,0.49,0.74}
\usepackage{hyperref}
\hypersetup{breaklinks,colorlinks,citecolor=cvprblue}

\def\paperID{*****} 
\def\confName{CVPR}
\def\confYear{2024}

\title{All-frequency Full-body Human Image Relighting}

\author{
Daichi Tajima \qquad Yoshihiro Kanamori \qquad Yuki Endo\\
University of Tsukuba, Japan\\
{\tt\small tajima.daichi.sp@alumni.tsukuba.ac.jp \qquad \{kanamori, endo\}@cs.tsukuba.ac.jp}
}

\begin{document}

\twocolumn[{
\renewcommand\twocolumn[1][]{#1}
\maketitle
\begin{center}
    \centering
    \captionsetup{type=figure}
    \includegraphics[width=\linewidth]{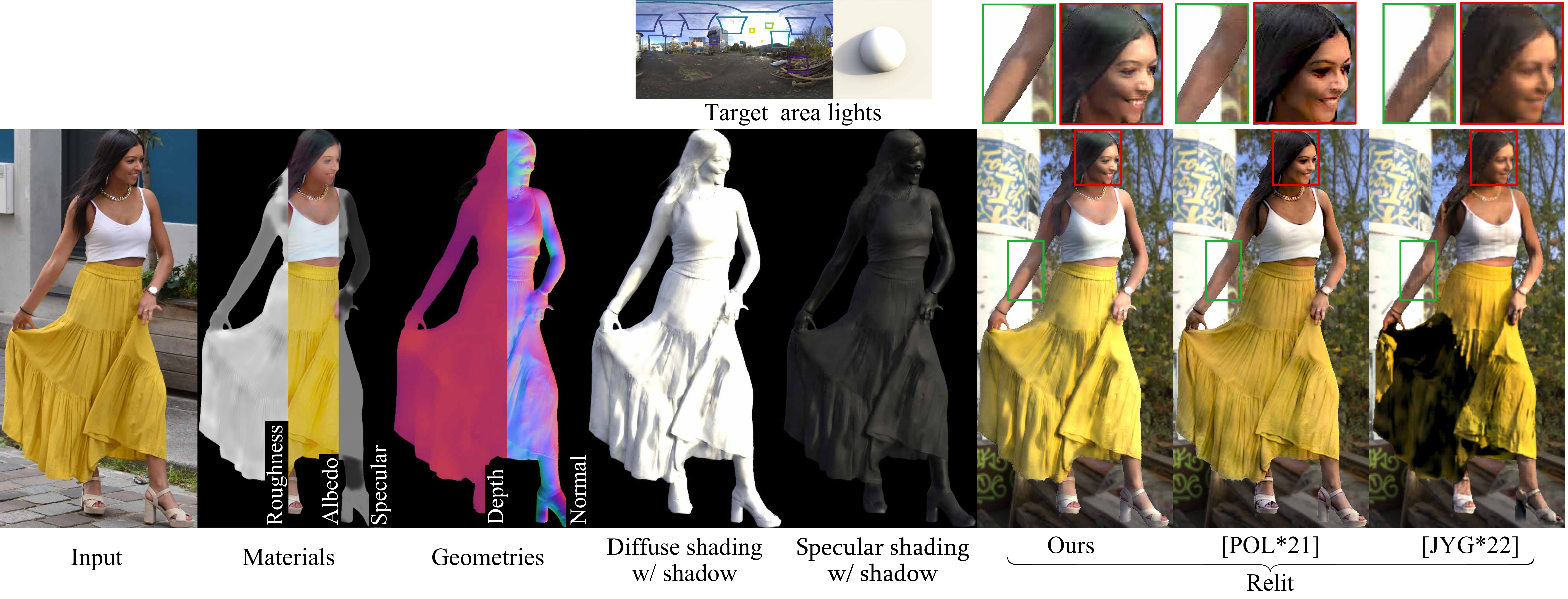}
    \captionof{figure}{
    Our method infers the materials (roughness, diffuse albedo, and specular) and geometry (depth and normal) from an input human image and calculates all-frequency shadows and reflections under new lighting conditions. As a lighting representation, we adopt a fixed number of area lights that approximate the target environment map.
    }
    \label{fig:teaser}
\end{center}
}]

\begin{abstract}
Relighting of human images enables post-photography editing of lighting effects in portraits. The current mainstream approach uses neural networks to approximate lighting effects without explicitly accounting for the principle of physical shading.
As a result, it often has difficulty representing high-frequency shadows and shading.
In this paper, we propose a two-stage relighting method that can reproduce 
physically-based shadows and shading from low to high frequencies.
The key idea is to approximate an environment light source with a set of a fixed number of area light sources. 
The first stage employs supervised inverse rendering from a single image 
using neural networks and calculates physically-based shading. The second stage then calculates shadow for each area light and sums up to render the final image.
We propose to make soft shadow mapping differentiable for the area-light approximation of environment lighting.
We demonstrate that our method can plausibly reproduce all-frequency shadows and shading caused by environment illumination, which have been difficult to reproduce using existing methods.
\end{abstract}  

\section{Introduction}
Human image relighting can alter the lighting effects in a portrait by changing the lighting condition after the photo shoot. The fundamental procedure for human image relighting is to infer the intrinsic geometry and reflectance of the target person as well as the scene illumination from the input image via inverse rendering and then render an output image with a new lighting condition. Modern learning-based methods formulate these inverse and forward rendering stages as a unified differentiable pipeline within an analysis-by-synthesis framework.

The current state-of-the-art techniques~\cite{SIG19,ICCV19_zhou,SIGA20_wang,CVPR20_nestmeyer,SIG21_Pandey,SIGA22} employ neural networks to approximate the forward rendering stage without explicitly considering the physical principles involved. In particular, the physical principle of shadows is often ignored; shadows appear when the target geometry occludes the incoming light. Explicitly modeling such light occlusion within a differentiable rendering pipeline has been proven challenging; recent approaches only support hard shadows caused by a single point/directional light~\cite{CVPR22,CVPR23} or adopt a computationally expensive solution via non-differentiable ray tracing with a pre-inferred geometry~\cite{ECCV22}. Consequently, neural networks in the state-of-the-art techniques struggle to learn complicated shadow patterns and yield blurry shadows or flickering artifacts with dynamic lighting.

In this paper, we step forward to reproduce physically plausible shadows for all-frequency relighting of human images. We simultaneously model the target geometry and environment illumination as a depth map and a fixed number of area lights within a differentiable framework to reproduce hard-to-soft shadows caused by multiple area lights. The ground-truth area lights for supervised learning are obtained via a novel optimization-based approach. We also infer the diffuse and specular reflectances of the target person for physically based shading.
Such geometry and reflectance information is easier to learn with neural networks because it is simpler than the complicated shadow and reflection patterns.
We demonstrate that our physically based formulation yields more plausible and stable relighting results even under dynamic lighting than the existing approximate solutions using neural networks (Figure~\ref{fig:teaser}).
We will release our source codes, trained models, and synthetic dataset upon publication.

\section{Related Work}
\label{sec:RelatedWork}
There have been numerous studies of single-image relighting. In the following, we focus on our main target, human image relighting. We categorize the previous work in terms of whether they explicitly calculate physically-based shading, use neural network approximations, or explicitly consider shape when calculating shadows.

\subsection{Relighting with Physically-based Shading}

To calculate physically-based shading, second-order spherical harmonics (SH) have often been used to account for diffuse-only environmental lighting.
MoFA~\cite{ICCV17} leverages morphable 3D face models~\cite{IEEE09} for face relighting.
SfSNet~\cite{CVPR18_sfs} employs face inverse rendering to estimate the normal map, diffuse reflectance, and illumination and then performs relighting by replacing the estimated illumination with a new illumination. 
These face relighting techniques ignore light occlusion due to the almost convex face shapes. In full-body relighting, however, light occlusion is common around limbs and cloth wrinkles and thus should not be ignored. As the first single-image full-body relighting method, Kanamori and Endo~\cite{SIGA18} extended SfSNet~\cite{CVPR18_sfs} to consider light occlusion explicitly in the second-order SH formulation by inferring light transport maps with light occlusion, instead of normal maps.

The diffuse-only method~\cite{SIGA18} is extended to handle specular reflections. Tajima \etal~\cite{PG21} introduced a refinement network module on top of \cite{SIGA18} to handle specular reflection and domain adaptation to in-the-wild photographs. Lagunas \etal~\cite{EGSR21} represented the per-pixel exiting radiance as a double product of fourth-order SH to account for various lighting effects, including specular reflection. However, it is well known that the low-order SH representations cannot represent high-frequency shading and shadows. Our method calculates all-frequency shading and shadows by each of multiple area lights without using SH representations.

\subsection{Relighting with Neural Network Approximations}

The traditional physically-based calculation of shading and shadowing is complicated and thus often fully or partially replaced with neural network approximations in modern relighting techniques. Early attempts of full approximations~\cite{SIG19,ICCV19_zhou} formulate relighting as image-to-image translation using single U-Net-like architectures, where new light information is injected at the bottleneck. Unfortunately, these methods do not have mechanisms to handle high-frequency signals. Song \etal~\cite{CGF21} proposed a relighting method for half-body portraits but requires another portrait as a reference of novel illumination.

Partial neural approximations calculate intermediate components and feed them into neural networks for final outputs. Pandey \etal~\cite{SIG21_Pandey} calculate multiple frequency bands of Phong-based specular components and merge them using a neural network. Yeh \etal~\cite{SIGA22} extended their work for domain adaptation in a similar spirit to Tajima \etal~\cite{PG21}. However, these methods do not consider light occlusion explicitly and thus cannot handle high-frequency shadows.
Yu \etal~\cite{ECCV20} used a neural network to estimate shadows during relighting of outdoor scenes.
Nestmeyer \etal~\cite{CVPR20_nestmeyer} and Wang \etal~\cite{SIGA20_wang} introduced neural network modules to estimate specular reflection and shadows. However, because these neural networks are unaware of the underlying geometry, they struggle to reproduce complicated patterns of specular reflection and shadows on diverse full-body human images, resulting in blurry shadings and shadows as well as flickering artifacts with dynamic lighting.

\begin{figure*}[t]
\centering
\includegraphics[width=\linewidth]{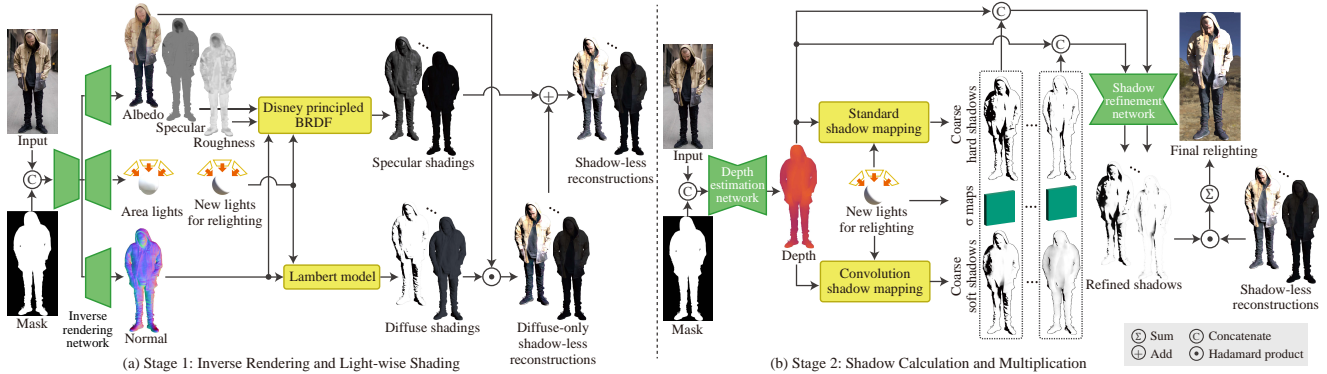}
\caption{
Overview of our two-stage relighting approach. Stage~1 applies inverse rendering and calculates a shadow-less shading image for each light. Stage~2 calculates shadows and multiplies shading images by the shadows and then merges them to output the relighting result.
}
\label{fig:flow}
\end{figure*}

\subsection{Geometry-aware Shadow Calculation}

Explicit calculation of light occlusion has been proven to be a key for high-frequency and stable shadows~\cite{EG22,CVPR23_sheng}. Ji \etal~\cite{ECCV22} reconstructs a 3D human model~\cite{CVPR20_saito} and applies ray tracing to calculate all-frequency shadows. Other techniques for reconstructing animatable 3D avatars~\cite{CoRR22_Corona,ICCV23} can be used in the same way. However, these methods support diffuse reflection only and rely on offline ray tracing, which is non-differentiable and computationally expensive. Hou \etal~\cite{CVPR21} estimated shadow regions on faces via 3D morphable model fitting but did not support environmental lighting. 

Differentiable shadow calculations with geometry information have appeared recently. Hou \etal~\cite{CVPR22} calculated visibility sampling for hard shadows by a directional light using ray marching with face depth maps. To extend this method for full-body images, however, we have to increase visibility samples due to the much more complicated geometry, which increases the computational burden. Even worse, this method entails incomplete shadows because depth maps have depth gaps. Worchel and Alexa~\cite{CVPR23} proposed a differentiable shadow mapping algorithm for a point/directional light source. They use variance shadow mapping (VSM)~\cite{SI3D06} to enable soft rasterization for differentiable calculation. Although VSM was originally proposed to calculate soft shadows, their method is tailored for point or directional light sources and thus cannot handle soft shadows as is. Contrarily, our method calculates differentiable soft shadows by an area light based on convolutional shadow mapping (CSM)~\cite{EGSR07}. Table~\ref{tab:ref_shadow} summarizes the taxonomy of recent techniques for geometry-aware shadow calculation.

\begin{table}[t] 
\caption{
Taxonomy of recent geometry-aware techniques for full-body human image relighting and shadow calculation. While \cite{CVPR20_nestmeyer,CVPR22} assume a single directional light, the inference times were measured using 16 lights similarly to ours.}
\label{tab:ref_shadow}
\hbox to\hsize{\hfil
\centering
\resizebox{\columnwidth}{!}{
\begin{tabular}{l|cccccc}
    \hline
                            & Approach                     & Light type                                  & \makecell{Geometry-\\aware}  & \makecell{Soft\\shadow}       & \makecell{Differen-\\tiable} & \makecell{Inference\\time (sec.)}\\
    \hline \hline
    \cite{CVPR20_nestmeyer} & CNN                          & \makecell{Directional}                & \redcross         & \redcross & \redcross & 0.351\\
    \cite{ECCV22}           & \makecell{Ray-tracing + CNN} & \makecell{Environmental}                  & \greencheck          & \greencheck & \redcross & 3.50\\
    \cite{CVPR22}           & Ray-marching             & \makecell{Directional}                & \greencheck & \redcross & \redcross & 12.5\\
    \cite{CVPR23}           & VSM                          & \makecell{Directional\&Spot}& \greencheck & \redcross & \greencheck & 0.150\\
    \hline
    Ours                    & \makecell{CSM + CNN}         & \makecell{Area lights}                   & \greencheck & \greencheck & \greencheck & 0.835\\ 
    \hline 
\end{tabular}\hfil}
}
\end{table}

\section{Method}
Figure~\ref{fig:flow} shows the overview of our method.
The inputs of our method are a full-body human image, its binary mask (obtained via off-the-shelf service~\cite{remove.bg} or software~\cite{photoshop}), and a new environmental illumination for relighting.
We employ a two-stage approach to handle shading and shadows as follows. In Stage~1, we first apply inverse rendering to obtain diffuse and specular reflectance, a set of area lights, and a normal map. We then calculate a shading image for each area light. In Stage~2, we estimate a depth map, calculate shadows for each area light, multiply the shadow maps pixel-wise by the shading images obtained in Stage~1, and then merge them to obtain the final output.

We approximate an environmental illumination as a set of a fixed number of area lights.
The motivation for using area lights as an approximation to HDRI maps is to explicitly capture the strong light that causes noticeable highlights and cast shadows, which is difficult to achieve with low resolution HDRI maps and lighting representations with low order spherical harmonics.
Let $N_L$ be the number of area lights. Each area light $l \in \{1, \dots, N_L\}$ is parameterized with an RGB intensity $\mathbf{L}^\mathit{int}_l \in \mathbb{R}^3$, light direction $\mathbf{L}^\mathit{dir}_l \in \mathbb{R}^3$, and area (denoted as the standard deviation $\sigma_l$ of light $l$'s Gaussian kernel). We represent these seven parameters for each area light as a light tensor $\mathbf{L} \in \mathbb{R}^{N_L\times7}$.

\subsection{Stage 1: Inverse Rendering and Light-wise Shading}
\label{subsec:1st}

In Stage~1, we perform inverse rendering and obtain a shading image lit by each area light while accounting for both diffuse and specular reflections.
First, we estimate a set of a fixed number of area light sources, normal maps, diffuse albedo, specular and roughness maps from the input image through the inverse rendering network, which has a U-Net-like architecture with a single encoder and three decoders.
The diffuse albedo and specular/roughness maps are estimated simultaneously by the same decoder.
Next, we calculate diffuse and specular shading. To simplify the shading calculation, we approximately handle each area light as a directional light; \ie, we ignore the area information, which is later utilized in calculating soft shadows in Stage~2.
The diffuse shading is calculated from the inferred normal map and each light based on the Lambert model.
The specular shading is calculated from the inferred normal map, specular/roughness maps and each light based on the Disney principled BRDF~\cite{SIG12}.
Note that these shading images do not contain shadows, which are later calculated and multiplied pixel-wise in Stage~2.

We organize the mathematical symbols used in this paper. The super-scripts ``$\mathit{diff}$'' and ``$\mathit{spec}$'' indicate diffuse and specular components, respectively. Hat {\textasciicircum} indicates inferred data. $l$-indexing indicates that the datum is inferred for area light $l$. Hat-less symbols are the corresponding ground truth, which are obtained using an offline ray tracer with an environment light (\ie, without discrete approximation and thus without $l$-indexing). We define inferred tensors as follows. Let $\hat{\mathbf{S}}_l$ and $\hat{\mathbf{V}}_l$ be shading and shadow images, respectively. We then define a \textit{shadowed shading} as $\hat{\mathbf{Y}}_l = \hat{\mathbf{V}}_l \odot \hat{\mathbf{S}}_l$ (where $\odot$ denotes the Hadamard product). We also define \textit{shadow-less reconstructions} for diffuse component $\hat{\mathbf{R}}^\mathit{diff}_l = \hat{\mathbf{A}} \odot \hat{\mathbf{S}}^\mathit{diff}_l$ (where $\hat{\mathbf{A}}$ denotes a diffuse albedo) and for both diffuse and specular components $\hat{\mathbf{R}}^\mathit{full}_l = \hat{\mathbf{A}} \odot \hat{\mathbf{S}}^\mathit{diff}_l + \hat{\mathbf{S}}^\mathit{spec}_l$.

For the supervised learning in Stage~1, we use L1 losses for the inferred diffuse albedo and specular/roughness/normal maps obtained via inverse rendering and for shading images $\{\hat{\mathbf{S}}_l\}$ and shadow-less reconstructions $\{\hat{\mathbf{R}}_l\}$ as follows.
\begin{align}
\label{eq:loss_decom_l1}
\mathcal{L}^\mathit{texture} &= \sum_{\mathbf{T} \in \mathcal{T}} \norm{\mathbf{T}-\hat{\mathbf{T}}}_1, \\
\mathcal{L}^\mathit{diff\,shading} &= \norm{\mathbf{S}^\mathit{diff}-\sum^{N_L}_{l=1}\hat{\mathbf{S}}^\mathit{\,diff}_l}_1, \label{eq:Diffhading} \\
\mathcal{L}^\mathit{spec\,shading} &= \norm{\mathbf{S}^\mathit{spec}-\sum^{N_L}_{l=1}\hat{\mathbf{S}}^\mathit{\,spec}_l}_1, \label{eq:SpecShading} \\
\mathcal{L}^\mathit{diff\,recon} &= \norm{\mathbf{R}^\mathit{diff}-\sum^{N_L}_{l=1}\hat{\mathbf{R}}^\mathit{diff}_l}_1, \label{eq:DiffRecons} \\
\mathcal{L}^\mathit{full\,recon} &= \norm{\mathbf{R}^\mathit{full}-\sum^{N_L}_{l=1}\hat{\mathbf{R}}^\mathit{full}_l}_1, \label{eq:FullRecons}
\end{align}
where $\mathcal{T}$ is a set of four types of texture maps $\mathbf{T}$, \ie, diffuse albedo, roughness map, specular map, and normal map.

We also use the VGG loss~\cite{CVPR18} for the diffuse albedo $\hat{\mathbf{A}}$, specular shading $\hat{\mathbf{S}}^\mathit{spec}_l$, and shadow-less reconstruction $\hat{\mathbf{R}}^\mathit{full}_l$:
\begin{align}
\label{eq:loss_decom_vgg}
\begin{split}
\mathcal{L}^\mathit{vgg} &= \textsc{vgg}(\mathbf{A},\hat{\mathbf{A}}) + \textsc{vgg}(\mathbf{S}^\mathit{spec},\sum^{N_L}_{l=1}\hat{\mathbf{S}}^\mathit{spec}_l) \\[-0.5em]
&+ \textsc{vgg}(\mathbf{R}^\mathit{full},\sum^{N_L}_{l=1}\hat{\mathbf{R}}^\mathit{full}_l),
\end{split}
\end{align}

\noindent
where \textsc{vgg} is a function to calculate the VGG loss.

The illumination of the input image is estimated so that input image can be reconstructed.
To learn the light parameters $\{\hat{\mathbf{L}}^\mathit{int}_l, \hat{\mathbf{L}}^\mathit{dir}_l, \hat{\sigma}_l\}$ for area light $l$, we do not use loss functions with their ground truth; our area light set is a discrete approximation of an environment light and is not necessarily unique. For example, there are countless arrangements of area lights to approximate a cloudy sky. In fact, we could not learn these parameters using loss functions with their ground truth. We instead learn these parameters via shading images $\{\hat{\mathbf{S}}_l\}$ and shadow-less reconstructions $\{\hat{\mathbf{R}}_l\}$. The light intensity $\hat{\mathbf{L}}^\mathit{int}_l$ and direction $\hat{\mathbf{L}}^\mathit{dir}_l$ are learned via Equations~(\ref{eq:Diffhading}) to (\ref{eq:FullRecons}). To learn the area information $\hat{\sigma}_l$, we calculate soft shadows using our differentiable convolutional shadow mapping (DCSM) function (elaborated in Section~\ref{subsub:shadow_mapping}), and compare with the ground-truth shadowed shading $\mathbf{Y}^\mathit{diff}$:
\begin{linenomath}
\begin{gather}
\tilde{\mathbf{V}}_l = \textsc{DCSM}(\mathbf{D},\mathscr{D}(\hat{\mathbf{L}}^\mathit{dir}_l),\hat{\sigma}_l), \\
\tilde{\mathbf{S}}^\mathit{diff}_l = \textsc{Lambert}(\mathbf{N},\mathscr{D}(\hat{\mathbf{L}}^\mathit{dir}_l),\mathscr{D}(\hat{\mathbf{L}}^\mathit{int}_l)), \\
\mathcal{L}^{\sigma} = \norm{\mathbf{Y}^\mathit{diff} - \sum^{N_L}_{l=1} \tilde{\mathbf{V}}_l \odot \tilde{\mathbf{S}}^\mathit{diff}_l}_1, \label{eq:LossSigma}
\end{gather}
\end{linenomath}
where \textsc{DCSM} and \textsc{Lambert} are functions to calculate soft shadows and Lambert shading, respectively. $\mathbf{D}$ and $\mathbf{N}$ denote the ground-truth depth and normal maps. $\mathscr{D}$ is a detaching operator to detach the argument's gradient from the computational graph. We detach the light intensity $\hat{\mathbf{L}}^\mathit{int}_l$ and direction $\hat{\mathbf{L}}^\mathit{dir}_l$ so that loss function $\mathcal{L}^{\sigma}$ can focus on the learning of $\sigma_l$; without detaching, we could not learn $\sigma_l$ well because of the ambiguity of soft shadows.

In summary, the final loss function used for Stage~1 is as follows:
\begin{align}
\begin{split}
\mathcal{L}_\mathit{decomposition} &= \mathcal{L}^\mathit{texture} + \mathcal{L}^\mathit{diff\,shading} + \mathcal{L}^\mathit{spec\,shading} \\
& +\mathcal{L}^\mathit{diff\,recon} + \mathcal{L}^\mathit{full\,recon} + \lambda_\mathit{vgg} \, \mathcal{L}^\mathit{vgg} + \lambda_{\sigma} \, \mathcal{L}^{\sigma},
\end{split}
\end{align}
where $\lambda_\mathit{vgg}=0.1$ and $\lambda_{\sigma}=0.01$.

\subsubsection{Background-aware light estmation}
\begin{figure}[t]
\centering
\includegraphics[width=1\linewidth]{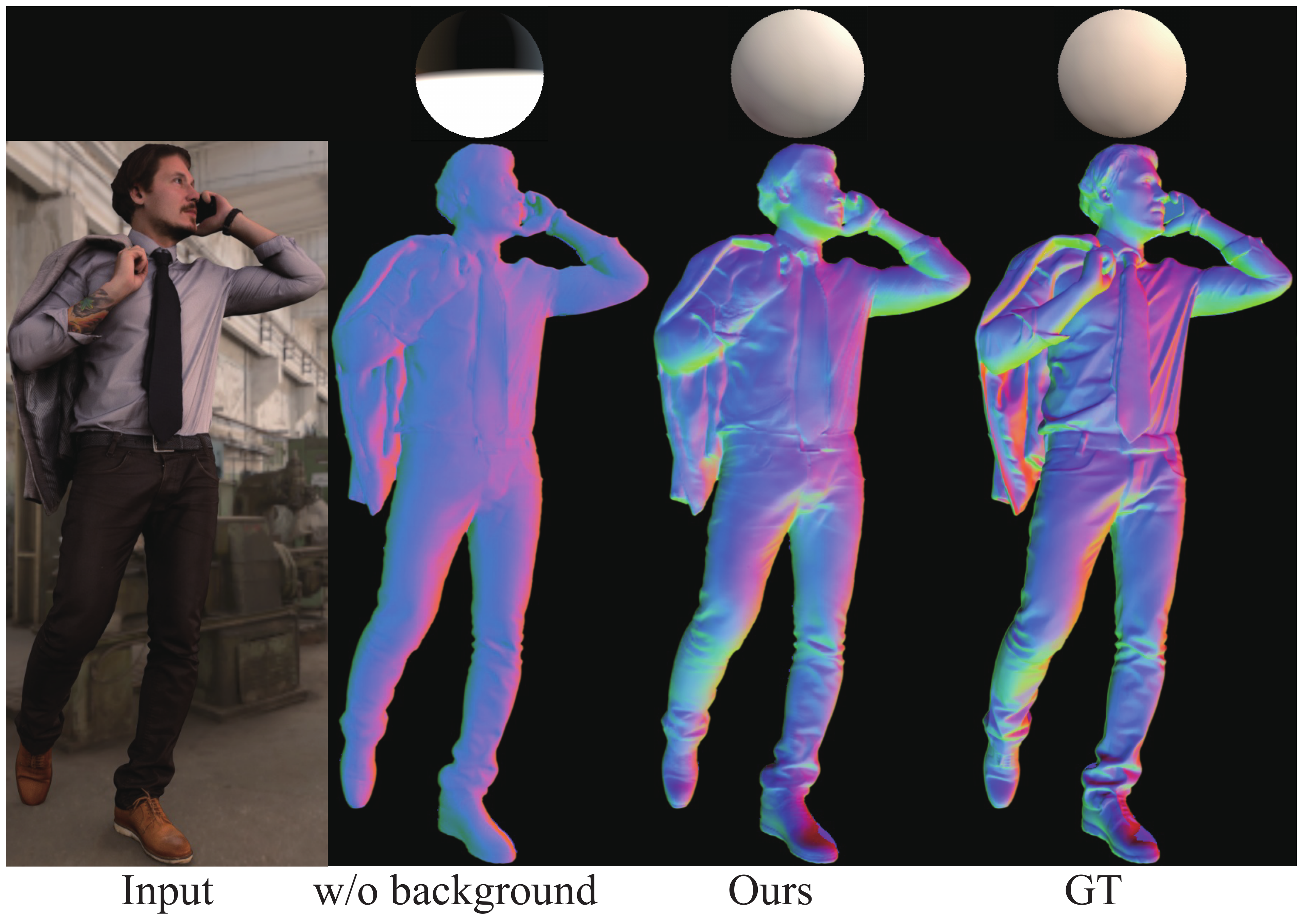}
\caption{
Validation of background-aware light estimation. Top row: spheres shaded with the estimated lights. Bottom row: estimated normal maps.
}
\label{fig:net_1st_bg}
\end{figure}

For light estimation, although previous methods~\cite{SIGA18,PG21,EGSR21} discard the background information in input images by multiplying them by binary masks, we exploit the background information to improve the light estimation accuracy. 
Specifically, we concatenate an input image (including the background) and a binary mask to feed the network.
Figure~\ref{fig:net_1st_bg} shows a qualitative comparison of the light source estimation with and without the background.
We can see that not only the inferred lights but also the inferred normal map is improved because the normal map is used together with lights to reproduce the shading in the input image.

\subsection{Stage~2: Shadow Calculation and Multiplication}
\label{subsec:2nd}
The goal of Stage~2 is to calculate shadows, multiply a shading image by the shadow for each light, and then merge the shadowed reconstructions to generate the final relighting result (see Figure~\ref{fig:flow}, right).
We first estimate a depth map from the input image using a depth estimation network.
Next, from the estimated depth map and area lights, we calculate shadows. Although we calculate soft shadows using our differentiable version of convolutional shadow mapping~\cite{EGSR07}, we find that the soft shadows do not exhibit sharp boundaries near the occludee's surface, which are crucial to reproduce all-frequency shadows. We thus calculate standard shadow mapping as well and merge the hard and soft shadows via a shadow refinement network. We elaborate on each step as follows.

\subsubsection{Depth estimation network}

We use a U-Net-like architecture to estimate a depth map from an input image (see Appendix~\ref{appendix:DepthEstimationNetwork} for our alternative network designs). 
To ignore absolute depth differences during training, we employ a scale-invariant L1 loss $\mathcal{L}_\mathit{si}$.
\begin{gather}
\mu_D = \frac{1}{\norm{\mathbf{M}}_1} \mathbf{M} \odot (\mathbf{D}-\mathscr{D}(\hat{\mathbf{D}}))
\label{eq:loss_depth_sil1}, \\
\mathcal{L}_\mathit{si} = \norm{\mathbf{D}-(\hat{\mathbf{D}}+\mu_D \, \mathbf{M})}_1, \label{eq:ScaleInvariant}
\end{gather}
where $\mathbf{M}$ is a binary mask and
$\mu_D$ is a scalar value.
While Eigen \etal~\cite{NIPS14_eigen} defined a scale-invariant loss as L2 loss with log-space depth because they wanted to emphasize near pixels while marginalizing far pixels. We do not have to use log-space depth because the depth of human bodies is within a short range, unlike general indoor/outdoor depth maps. In our case, however, we cannot determine the absolute depth only with the scale-invariant loss, so we restrict output depth values within $[0, 1]$ by adding a sigmoid function at the last layer in the network.

We also introduce a regularization term to smooth the surface because slight irregularities in the depth map can cause serious artifacts in the shadow calculation.
Specifically, we apply L1 loss to the gradient of depth maps, as Eigen and Fergus~\cite{ICCV15_eigen} did with L2 loss:
\begin{align}
\label{eq:loss_depth_slope}
\mathcal{L}_\mathit{slope} = \norm{\frac{\partial \mathbf{D}}{\partial x}-\frac{\partial \hat{\mathbf{D}}}{\partial x}}_1 + \norm{\frac{\partial \mathbf{D}}{\partial y}-\frac{\partial \hat{\mathbf{D}}}{\partial y}}_1.
\end{align}
The final loss function is
\begin{equation}
\mathcal{L}_\mathit{depth} = \mathcal{L}_\mathit{si} + \lambda_\mathit{slope} \, \mathcal{L}_\mathit{slope},
\label{eq:loss_depth}
\end{equation}
where $\lambda_\mathit{slope}=0.01$.

\subsubsection{Differentiable convolutional shadow mapping (DCSM)}
\label{subsub:shadow_mapping}

While the original convolutional shadow mapping (CSM)~\cite{EGSR07} is non-differentiable, we make it differentiable and implement it using nvdiffrast~\cite{SIGA20_laine}. With our differentiable CSM (or DCSM), we can learn the area information of area lights required for calculating soft shadows (Section~\ref{subsec:1st}) and obtain a set of area lights as a discrete approximation of an environmental illumination via optimization (Section~\ref{sec:arealightdataset}).

We briefly review CSM.
Let $\mathbf{x}$ be a point on the surface visible to the camera, $d(\mathbf{x})$ the distance from $\mathbf{x}$ to the light source, $\mathbf{p}$ the position of the obstacle when trying to view $\mathbf{x}$ from the light source, and $z(\mathbf{p})$ the distance from $\mathbf{p}$ to the light source.
The binary shadow test function, which determines whether a point is in shadow or not, is defined as follows:
\begin{equation}
f(d(\mathbf{x}),z(\mathbf{p}))=
\begin{cases}
1 &  \text{if} \,\,\,\, d(\mathbf{x}) \leq z(\mathbf{p})\\
0 &  \text{otherwise,}
\end{cases}
\label{eq:csm_shadowtestfunction}
\end{equation}
where 0 indicates shadowed and 1 unshadowed.
$f$ is essentially a Heaviside step function and therefore discontinuous. CSM approximates Equation(\ref{eq:csm_shadowtestfunction}) with a continuous function by expanding it with Fourier series.
\begin{align}
f(d(\mathbf{x}),z(\mathbf{p})) \approx \sum_{i=1}^K \mathbf{a}_i (d(\mathbf{x})) \, \mathbf{B}_i (z(\mathbf{p})),
\label{eq:csm_approximate}
\end{align}
where $B_i$ is a basis function of $z(\mathbf{p})$, each basis being weighted by a coefficient $a_i$ which depends on $d(\mathbf{x})$.
$K$ is the truncation order, and it is known that small $K$ reduces the approximation accuracy, causing light bleeding and ringing.
We set $K = 8$ in our method.
CSM can vary the shadow hardness/softness via convolution with an arbitrary kernel $w_{\sigma}$.
The convolved version $s_{f}$ of $f$ is as follows:
\begin{align}
\begin{split}
s_{f}(d(\mathbf{x}),z(\mathbf{p})) &= [\mathbf{w}_{\sigma} \ast \sum_{i=1}^K \mathbf{a}_i (d(\mathbf{x})) \, \mathbf{B}_i (z(\mathbf{p}))](\mathbf{p}) \\[-0.5em]
&= \sum_{i=1}^K \mathbf{a}_i (d(\mathbf{x})) [\mathbf{w}_{\sigma} \ast \mathbf{B}_i (z(\mathbf{p}))](\mathbf{p}),
\end{split} \label{eq:csm_conv}
\end{align}
where $[\mathbf{w} \ast \mathbf{g}](\mathbf{p})$ means a convolution of $\mathbf{g}$ by the kernel $\mathbf{w}$ in the neighborhood of $\mathbf{p}$.
In CSM, the shadow hardness is determined by the kernel size, which is an integer value and thus non-differentiable.

To make CSM differentiable, we control the kernel size indirectly via the standard deviation $\sigma$ of a Gaussian kernel, which is continuous and thus differentiable. We determine the kernel size as $2 \lceil 3 \sigma \rceil + 1$ (where $\lceil x \rceil$ denotes the smallest integer equal to or larger than $x$) because $6 \sigma$ has more than 99\% coverage.

As an alternative to CSM, we also made exponential shadow mapping (ESM)~\cite{GI08} differentiable because ESM was proposed as an improved version of CSM. However, we found such differentiable ESM slows down optimization; clamping values greater than one in the shadow test function hinders gradient propagation.

\subsubsection{Shadow refinement network}

Unfortunately, our DCSM inherits the limitation of the original CSM; although soft shadows in the real world have varying shadow hardness because the penumbra widths vary with the distance between the occluder and occludee,  CSM does not take this into account. Even worse, shadows obtained from a depth map are often incomplete due to the missing geometry between depth gaps.

To address these problems, we also calculate hard shadows for high-frequency shadows and merge the hard and soft shadows via a shadow refinement network that also has a U-Net-like architecture. The hard shadows are calculated from a 3D mesh reconstructed from the depth map as a byproduct of DCSM and thus without additional burden. We find that the standard deviation $\sigma$ of a Gaussian kernel well represents the shadow softness, so we feed not only the hard and soft shadows but also a constant map of $\sigma$ after concatenating them for each area light (see Section~\ref{subsec:ablationstudy_shadow} for the ablation study with and without $\sigma$). 

To train the shadow refinement network, we define a loss function with the refined shadow $\bar{\mathbf{V}}_l$ obtained from the shadow refinement network and the ground-truth shadowed shading $\mathbf{Y}^\mathit{diff}$ as follows:
\begin{gather}
\mathbf{S}^\mathit{diff}_l = \textsc{Lambert}(\mathbf{N},\mathbf{L}^\mathit{dir}_l,\mathbf{L}^\mathit{int}_l), \\
\mathcal{L}_\mathit{refinement} = \norm{\mathbf{Y}^\mathit{diff}-\sum_{l=1}^{N_L} \bar{\mathbf{V}}_l \odot \mathbf{S}^\mathit{diff}_l}_1.
\label{eq:loss_refineshadow}
\end{gather}

\begin{figure*}[t]
\centering
\includegraphics[width=1\linewidth]{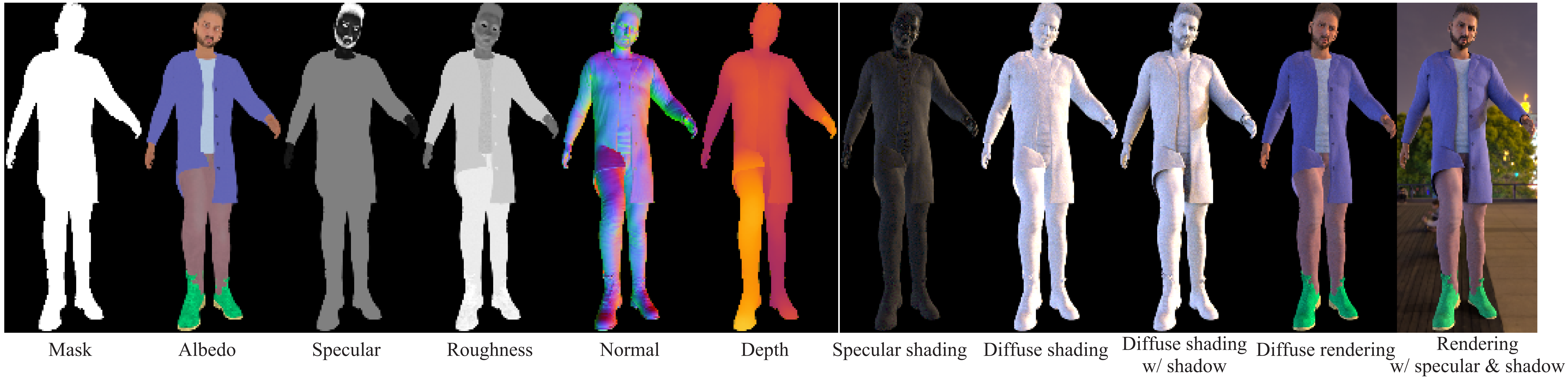}
\caption{
Example data generated from a 3D human model.
}
\label{fig:dataset_human}
\end{figure*}

\begin{figure}[t]
\centering
\includegraphics[width=1\linewidth]{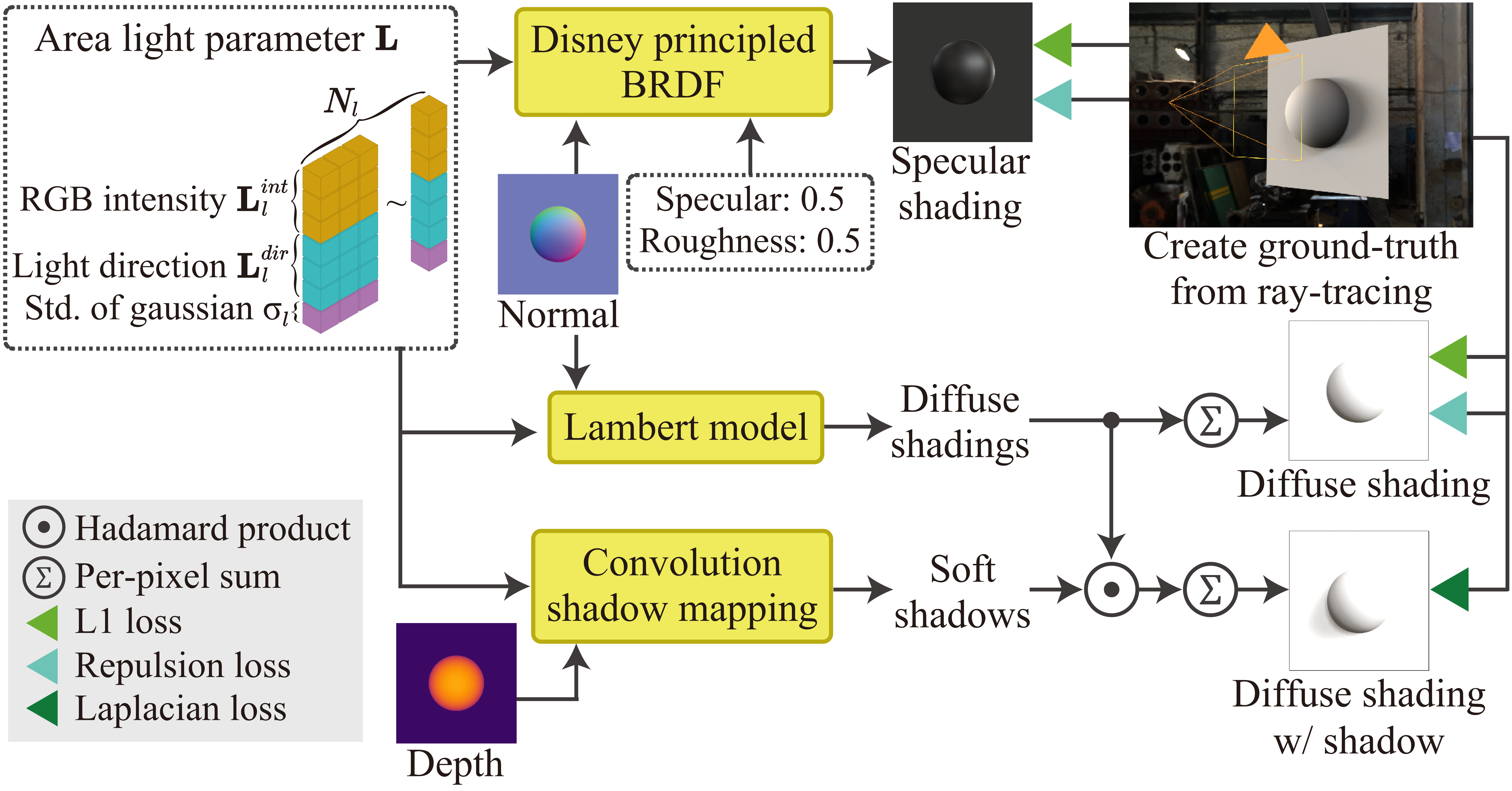}
\caption{
Overview of our area light optimization.
}
\label{fig:flow_opt_light}
\end{figure}

\begin{figure}[t]
\centering
\includegraphics[width=1\linewidth]{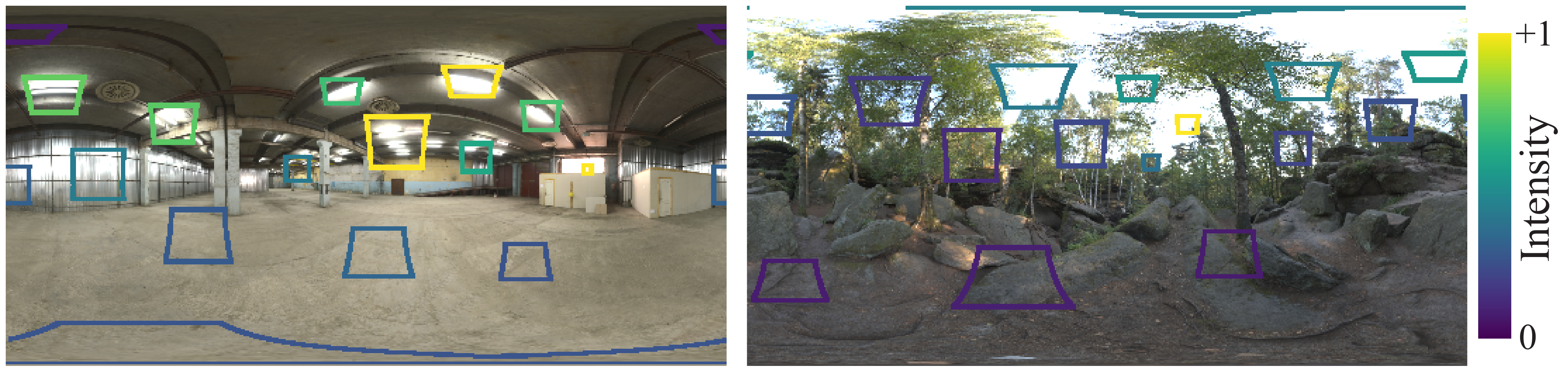}
\caption{
Visualization of optimized light parameters on an environment map. The light's area is visualized as the rectangle size, while the rectangle color indicates the light's intensity.
}
\label{fig:dataset_lightmapping}
\end{figure}

\section{Dataset}
\label{sec:dataset}

\subsection{Full-body Human Dataset}
\label{sec:humandataset}
We create a synthetic 3D human model dataset using Blender's add-on tool for generating 3D human models~\cite{hgv3} to obtain ground-truth full-body human image data.
The 3D human models include non-diffuse materials based on a simplified version of the Disney principled BRDF, but the subsurface scattering, anisotropy, and metal parameters are not considered.
The identity, standing pose, clothing, and camera direction of each 3D human model were randomly determined, resulting in 2,500 3D human models with different identities.
Of these, 2,400 were used for training and 100 for testing.
For each 3D human model, we render a binary mask, diffuse albedo map, specular map, roughness map, normal map, and depth map at a resolution of $1024\times1024$ pixels.
The ground truth data for the relighting images was rendered using ray tracing with a virtual light stage created in Blender using HDR environment maps collected from Poly Haven~\cite{polyhaven} without considering indirect illumination.
487 environment maps were used for training and 34 for testing. 
For each 3D human model, eight environment maps were randomly selected and randomly rotated along the longitude to increase the variation.
Figure~\ref{fig:dataset_human} shows some example data used in our experiments.
We also used 541 scanned 3D human models obtained from various commercial websites to further increase the variation.
Note that some of the scanned models do not contain ground-truth specular and roughness maps, so for such data, we omit the corresponding loss functions during training.

\subsection{Area Light Dataset}
\label{sec:arealightdataset}

We create a dataset of our area light approximations of environmental lights. The source environment maps are the same HDR images as those used for background images. To construct a discrete approximation of environmental illumination, Annen \etal~\cite{SIG08} proposed a greedy algorithm that outputs a varying number of area lights. Their approach is not suited for our purpose because we want a fixed-size tensor to learn using our inverse rendering network. We propose a novel optimization-based algorithm to output a fixed number of area lights.

As discussed on $\mathcal{L}^{\sigma}$ (Equation~\ref{eq:LossSigma}) in Section~\ref{subsec:1st}, we focus on shading images as the cue for optimization. Specifically, we put a hemisphere on a plane, illuminate the scene with the target environment illumination, and arrange area lights so that the shading and shadows match with those lit by the environment illumination. Not to miss strong incoming lights, we rotate the environment map five times by $72^\circ$ in longitude and $\pm{90}^\circ$ in latitude, resulting in seven images of shading and shadows rendered using ray tracing at the resolution of $256 \times 256$ pixels. Each optimization iteration accounts for these seven images simultaneously (for simplicity, we omit the loop for these seven images). Figure~\ref{fig:flow_opt_light} shows the overview of the optimization process. 
We optimize light intensity $\mathbf{L}^\mathit{int}_l$, direction $\mathbf{L}^\mathit{dir}_l$, and standard deviation $\sigma_l$ of light $l$'s Gaussian kernel for each area light $l \in \{1, 2, \dots, N_L\}$. The optimization process undergoes the following three steps:

\begin{enumerate}[Step 1:]
\item Initialize $\{\sigma_l,\mathbf{L}^\mathit{int}_l,\mathbf{L}^\mathit{dir}_l\}$,
\item Optimize $\mathbf{L}^\mathit{int}_l$ and $\mathbf{L}^\mathit{dir}_l$ of each light $l$, and
\item Optimize $\sigma_l$ of each light $l$, while fixing $\mathbf{L}^\mathit{int}_l$ and $\mathbf{L}^\mathit{dir}_l$.
\end{enumerate}

Step~1 initializes the light directions so that $N_L$ lights distribute uniformly on the environment map. The light intensities and $\sigma_l$ are initialized with constant values. The initial value of $\sigma_l$ depends on the shadow map resolution. In our case, we set $\sigma_l = 10$ for a $256 \times 256$ shadow map.

Step~2 utilizes both diffuse and specular shadings to optimize the intensity and direction of each light. As alternative strategies, diffuse-only reflection results in blurry shading and yields a high degree of freedom in light directions, whereas specular-only shading causes a concentration of light directions contributing to the most glossy directions, reducing the reproducibility of diffuse shading. We set the specular and roughness parameters as 0.5. The following L1 loss functions are used for optimization:
\begin{align}
\label{eq:loss_opt_shading}
\mathcal{L}_\mathit{diff} &= \norm{\mathbf{S}^\mathit{diff\,sph}-\hat{\mathbf{S}}^\mathit{diff\,sph}}_1, \\
\mathcal{L}_\mathit{spec} &=  \norm{\mathbf{S}^\mathit{spec\,sph}-\hat{\mathbf{S}}^\mathit{spec\,sph}}_1.
\end{align}
Furthermore, each light direction is encouraged to move away from each other to avoid overlapping light directions.
Specifically, we add a regularization term so that the dot product of each light direction pair is not greater than $\tau$.
\small
\begin{align}
\label{eq:loss_opt_rep}
\mathcal{L}_\mathit{rep} = \frac{1}{N_L(N_L-1)}\displaystyle \sum_{l=1}^{N_L-1} \sum_{m \ne l}^{N_L-1} \left| \text{max}(\tau,\left<\hat{\mathbf{L}}^\mathit{dir}_l,\hat{\mathbf{L}}^\mathit{dir}_m\right>-\tau) \right|^2 \mspace{-5mu} .
\end{align}
\normalsize
We set $\tau = 0.65$.

In Step~3, na\"ively optimizing $\sigma_l$ in terms of luminance does not work well because most pixels are inside or outside shadows and their luminance values do not change during optimization. We thus put emphasis on shadow boundaries by applying a Laplacian filter:
\small
\begin{gather}
\hat{\mathbf{V}}^\mathit{sph}_l = \text{DCSM}(\mathbf{D}^\mathit{sph},\mathscr{D}(\hat{\mathbf{L}}^\mathit{dir}_l),\hat{\sigma}_l), \\
\mathcal{L}_\mathit{lap} = \mspace{-30mu} \displaystyle \sum_{k \in \{15,21,33\}} \! \norm{\mathscr{L}_k \mspace{-2mu} \left( \! \mathbf{Y}^\mathit{diff\,sph}\right) \! - \! \mathscr{L}_k \mspace{-2mu} \left(\sum_{l=1}^{N_L} \hat{\mathbf{V}}^\mathit{sph}_l \mspace{-3mu} \odot \mathscr{D}(\hat{\mathbf{S}}^\mathit{diff\,sph}_l) \!\! \right)}_1 \mspace{-7mu} , \label{eq:loss_opt_shadow}
\end{gather}
\normalsize
\noindent
where $\mathbf{D}^\mathit{sph}$ is the ground-truth depth map, $\hat{\mathbf{V}}^\mathit{sph}$ is the inferred shadows of each light source, $\mathscr{L}_k$ is a Laplacian operator with kernel size $k\!\times\!k$, and $\mathbf{Y}^\mathit{diff\,sph}$ is the ground-truth shadowed shading for diffuse component.

Regarding the choice of the number of area lights, $N_L$, the larger $N_L$ is, the better accuracy we can obtain with more computational cost. We use $N_L = 16$ throughout this paper due to the trade-off between the accuracy and computational cost. See Appendix~\ref{appendix:NoOfLights} for our experiment with different $N_L$.

The optimization was implemented using Python and PyTorch and performed on NVIDIA RTX A5000.
We used Adam as an optimizer, setting the exponential decay rates for the moment estimates as \{0.5, 0.999\}.
The learning rate was controlled in the range of [1, 0.00001] by the cosine annealing scheduler within 20 epochs per cycle.
Each step took 1,000 iterations and 300 iterations to converge. The optimization took about 5 minutes per environment map. Figure~\ref{fig:dataset_lightmapping} visualizes the final area light parameters. For each light, the light intensity is color-coded and $\sigma_l$ value is visualized as the size of rectangle.
We can observe that the area lights are distributed appropriately in the environment map.

\section{Experiments}

\subsection{Implementation details}
We implemented our method using Python and PyTorch and performed training and inference on NVIDIA RTX A5000. 
We trained the inverse rendering network (Stage~1),
depth estimation network, shadow refinement network (Stage~2) separately. We used Adam as an optimizer, setting the exponential decay rates for the moment estimates as \{0.5, 0.999\}. 
We used the cosine annealing scheduler to control the learning rate in the range of [0.01, 0.00001] within 20 epochs per cycle.
Our batch size was eight. 
The computational times for one-epoch training of the inverse rendering network,
depth estimation network, and shadow refinement network were about 60, 50, and 50 minutes, respectively, when we used one GPU to process 1024 × 1024 images. We terminated the training at 300, 500, and 120 epochs for the respective networks, where each learning curve reached a plateau. The time for testing a $1024 \times 1024$ input image was about 1.01 seconds.
As the time breakdown, inverse rendering network,
depth estimation network, shadow mapping, shadow refinement network, and relighting took about 0.156, 0.0322, 0.747, 0.0875, and 0.0316 seconds, respectively.

\subsection{Ablation Studies}
We first conduct ablation studies using our dataset to evaluate the effectiveness of our method for improving shadows and specular components. 

\subsubsection{Comaprison with different types of shadows}
\label{subsec:ablationstudy_shadow}
\begin{figure*}[t]
\centering
\includegraphics[width=1\linewidth]{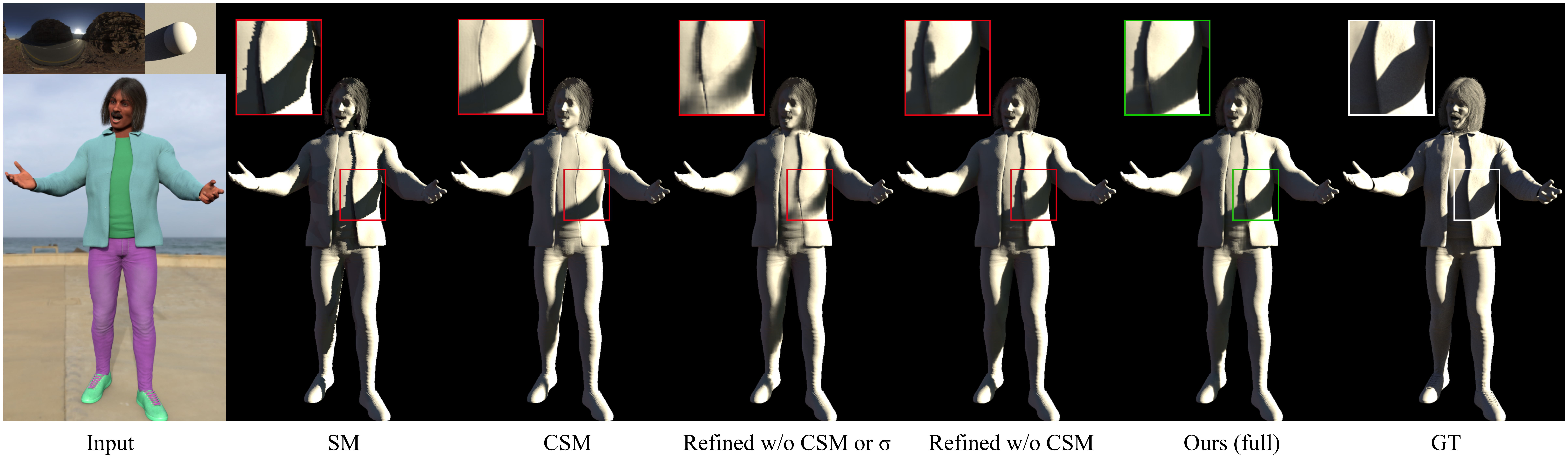}
\caption{
Ablation study of shading with shadows.
The environment map for relighting and its reference shading/shadowing pattern with a sphere are visualized above the input image.
The rectangles emphasize the differences. Each method is explained in Section~\ref{subsec:ablationstudy_shadow}. 
}
\label{fig:abl_diffuseshadingwshadow_cg}
\end{figure*}

Figure~\ref{fig:abl_diffuseshadingwshadow_cg} shows a comparison with standard shadow mapping (SM), convolutional shadow mapping (CSM), and our shadow refinement network. Here, ``Refined w/o CSM or $\sigma$'' means that the shadow refinement network does not use soft shadows by CSM or standard deviation $\sigma$ as input, and ``Refined w/o CSM'' means that it does not use soft shadows by CSM. 
As can be seen in the results, ``SM'' exhibits noticeable jaggies in the shadow contours by trying to approximate the environment light using a small number of lights. ``CSM'' can mitigate this problem using area lights but causes light bleeding when an occludee is near an occluder. In addition, ``CSM'' cannot reproduce contact shadows according to the distance between an occludee and an occluder. ``Refined w/o CSM or $\sigma$'' can reproduce contact shadows, but the shadows have the same softness regardless of the lights because no area light information is given. Meanwhile, ``Refined w/o CSM'', which uses a $\sigma$ map as input, reproduces not only contact shadows but also shadow softness depending on the area lights. Using soft shadows as an additional input, ``Ours (full)'' reproduces the most accurate shadows. 
In addition, as shown in the quantitative results in Table~\ref{tab:diffuseshadingwshadow_cg}, we can see step-by-step improvements by using CSM and a $\sigma$ map as input of ``Ours (full)''. Ours also achieved the best scores in all metrics. 

\subsubsection{Comparison of specular components}
\begin{figure}[t]
\centering
\includegraphics[width=1\linewidth]{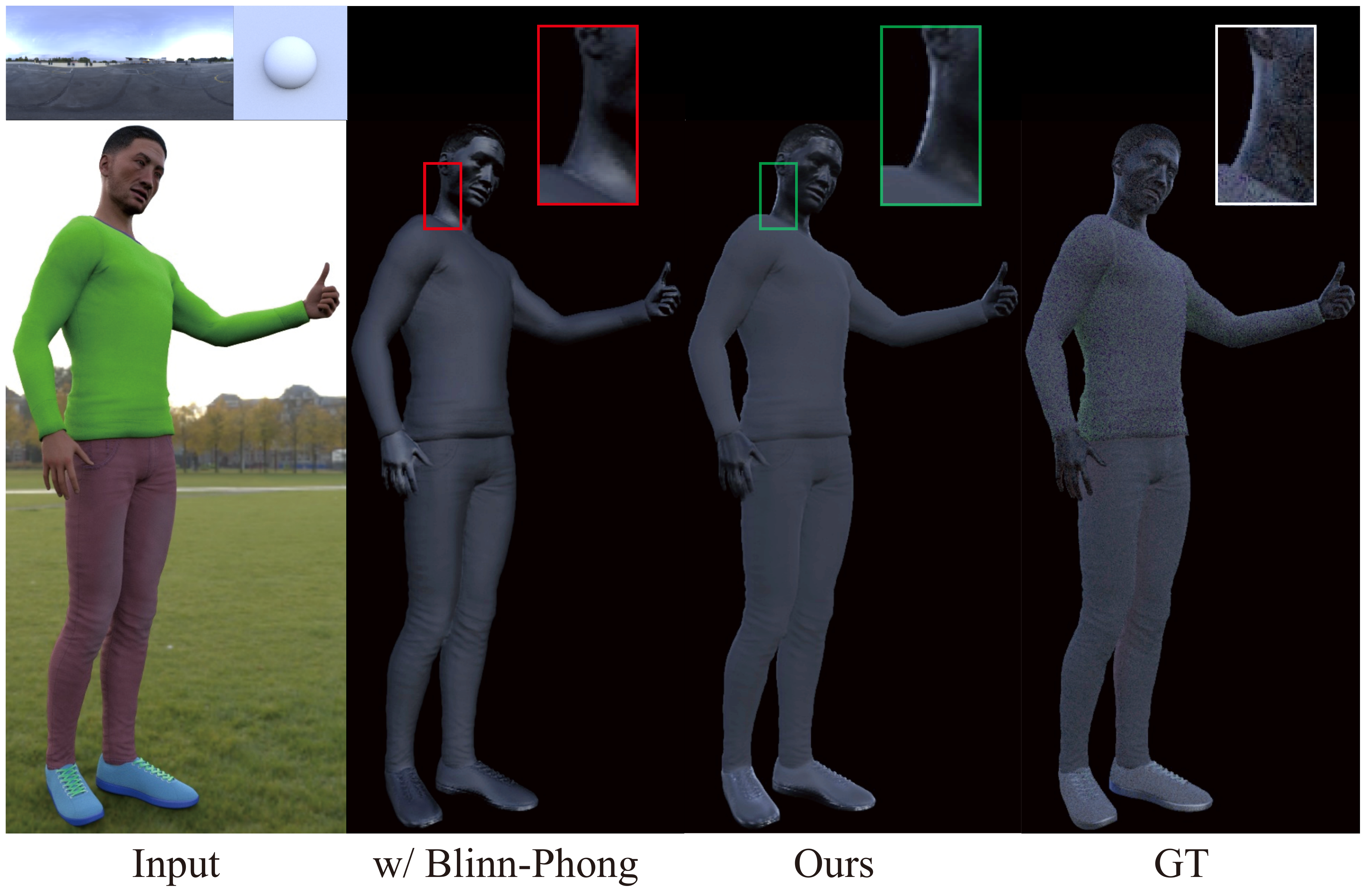}
\caption{
Ablation study of specular components.
For better visualization, the output images are uniformly multiplied by a constant factor.}
\label{fig:abl_specularshading}
\end{figure}

To evaluate the effectiveness of specular shading by the Disney principled BRDF, we compare it with the Blinn-Phong reflection model. To do so, we estimate a Blinn-Phong specular exponent map instead of a specular map and a roughness map using the inverse rendering network. Because ground-truth exponent maps are unavailable,
we trained the model by computing the losses only for relit images containing specular shading.
Figure~\ref{fig:abl_specularshading} and Table~\ref{tab:specularshading_cg} show the qualitative and quantitative results, respectively. The Blinn-Phong reflection model causes errors, especially around grazing angles, 
because it is less physically-plausible.
In addition, we can see that a single parameter is insufficient to reproduce reflectance properties in clothed full-body human images. 

\subsubsection{Comparison of relighting}
\label{subsub:comparison_relighting}
We verified the effectiveness of the shadows and specular reflections considered in our method with relighting results.
We quantitatively compared the following four conditions:
no shadow or specularity (``Ours w/o shadow or specular''), no shadow refinement or shadow (``Ours w/o refinement or specular''), no specular (``Ours w/o specular'') and ``Ours (full)''. A quantitative comparison of the relit results is shown in Table~\ref{tab:relit_cg}.
``Ours w/o refinement or specular'' with direct use of shadows by CSM improved accuracy in all metrics except LPIPS compared to ``Ours w/o shadow or specular''.
``Ours w/o specular'' with shadow refinement shows a further improvement in the accuracy of estimating the relighting results.
Furthermore, ``Ours (full)'', which takes specular into account, shows a significant effect on the accuracy improvement.

\subsection{Comparison with Existing Methods}
We compared our method with existing relighting methods specialized for human face images~\cite{SIG19,CVPR20_nestmeyer,CVPR22}, upper body images~\cite{SIG21_Pandey}, and full-body humans~\cite{EGSR21,PG21,ECCV22}. We used the public pre-trained models for several methods~\cite{PG21,EGSR21} and trained the models using our CG dataset from scratch for the other methods. For the methods~\cite{CVPR20_nestmeyer,CVPR22}, which assume relighting with a single directional light, 
we used 16 directional lights obtained from our 16 area lights, while ignoring $\sigma$, for a fair comparison. 

\subsubsection{Comparison of shadows}
\label{subsub:comparison_shadow}
\begin{figure}[t]
\centering
\includegraphics[width=1\linewidth]{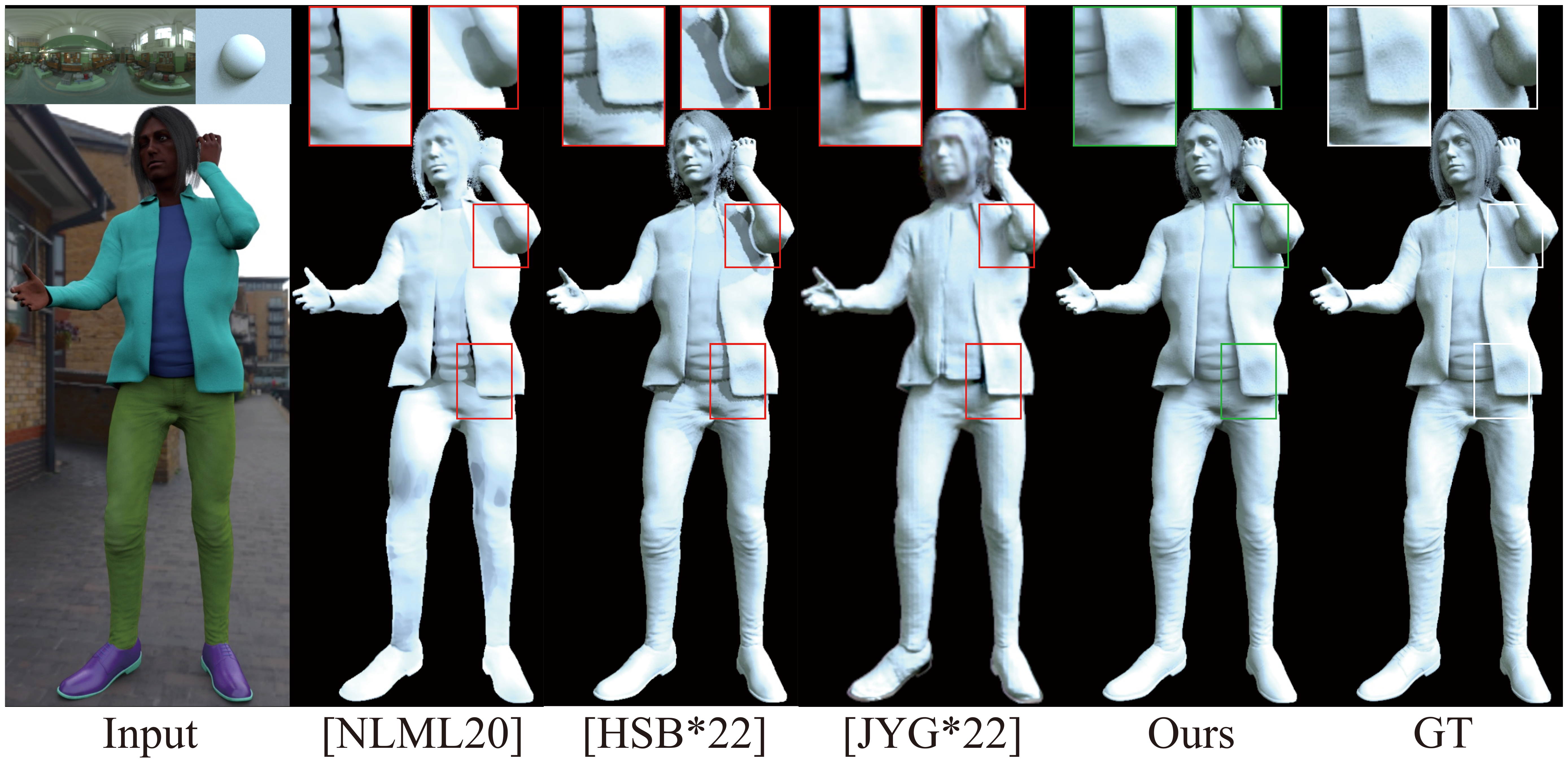}
\caption{
Qualitative comparison of shading with shadows between our method and the existing methods.
}
\label{fig:diffuseshadingwshadow_cg}
\end{figure}

Table~\ref{tab:diffuseshadingwshadow_cg} and Figure~\ref{fig:diffuseshadingwshadow_cg} show the quantitative and qualitative comparisons with the existing relighting methods that explicitly generate shadows.
The method by Nestmeyer \etal~\cite{CVPR20_nestmeyer} fails to reproduce shadows because it does not consider human shapes during visibility estimation with a CNN.
This method also has the highest standard deviation compared to the other methods. 
This is because learning to estimate shadows with varying illumination and geometry
is more difficult than estimating shape alone and performing physical lighting calculations.
Although the method by Hou \etal~\cite{CVPR22}, like our method, uses meshes reconstructed from depth maps, it reproduces only hard shadows caused by directional lights and yields artifacts around the shadow boundaries. The method by Ji \etal~\cite{ECCV22} also cannot obtain satisfactory results because of incomplete meshes estimated using PIFuHD~\cite{CVPR20_saito} and insufficient refinement by their shadow refinement module. 
In contrast, our method has fewer shadow artifacts due to the lack of complete geometry in the invisible regions.
Furthermore, the use of area light sources allows us to reproduce low to high-frequency shadows.

As for computational time, these existing methods took about 17 seconds~\cite{CVPR22} and 13 seconds~\cite{ECCV22}
to relight a single image.
These computational times consist mainly of 3D reconstruction and shadow calculation. 
Meanwhile, our method is much faster and took about 1 second for inference.

\begin{table}[t] 
\caption{
Quantitative comparison of shading with shadows (mean$\pm$standard deviation). The best scores are in bold, and the second-best scores are underlined. 
Each method is explained in 
Section~\ref{subsec:ablationstudy_shadow}. 
} 
\label{tab:diffuseshadingwshadow_cg}
\hbox to\hsize{\hfil
\centering
\resizebox{\columnwidth}{!}{
\begin{tabular}{l|ccc}
    \hline
                         & RMSE$\downarrow$ & SSIM$\uparrow$  & LPIPS$\downarrow$\\
    \hline \hline
    \cite{CVPR20_nestmeyer}& 0.457\small{$\pm$0.627}      & 0.457\small{$\pm$0.190}     & 0.106\small{$\pm$0.0215}\\
    \cite{CVPR22}        & 0.141\small{$\pm$0.0590}      & 0.636\small{$\pm$0.114}    & 0.0722\small{$\pm$0.0197}\\
    \cite{ECCV22}        & 0.145\small{$\pm$0.0619} & 0.612\small{$\pm$0.110} & 0.0891\small{$\pm$0.0156}\\ 
    \hline
    Ours                 & \textbf{0.100}\small{$\pm$0.0454} & \textbf{0.709}\small{$\pm$0.110} & \textbf{0.0610}\small{$\pm$0.0180} \\ 
    \hline
    Refined w/o CSM      & \underline{0.101}\small{$\pm$0.0455} & \underline{0.708}\small{$\pm$0.110} & \underline{0.0611}\small{$\pm$0.0180}\\
    Refined w/o CSM or σ & 0.102\small{$\pm$0.0462} & 0.704\small{$\pm$0.109} & 0.0618\small{$\pm$0.0180}\\
    CSM                  & 0.142\small{$\pm$0.0548} & 0.652\small{$\pm$0.111} & 0.0703\small{$\pm$0.0203}\\
    SM                   & 0.135\small{$\pm$0.0542} & 0.646\small{$\pm$0.112} & 0.0702\small{$\pm$0.0190}\\
    \hline
\end{tabular}\hfil}
}
\end{table}

\subsubsection{Comparison of specular components}
\label{subsub:comparison_specular}

\begin{figure}[t]
\centering
\includegraphics[width=1\linewidth]{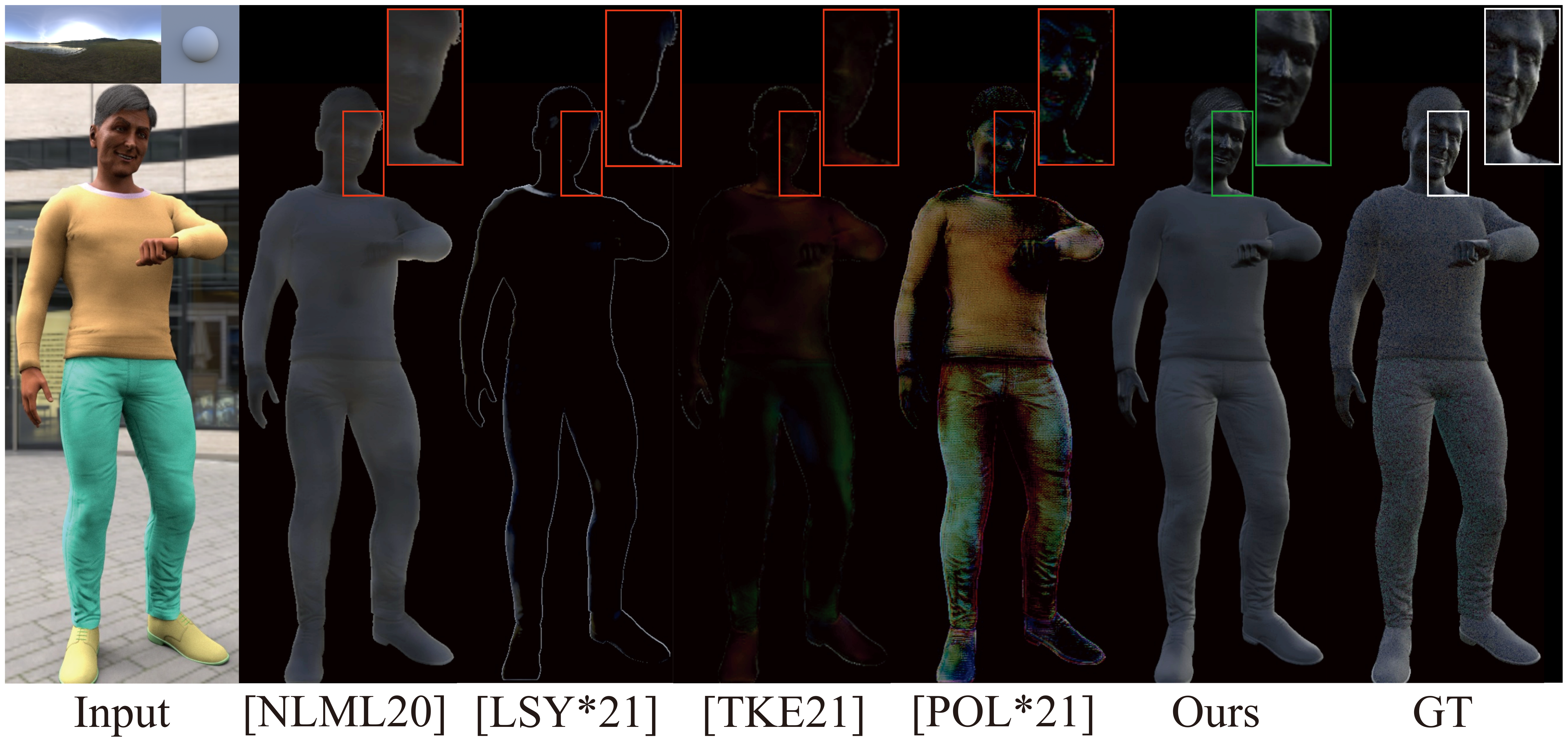}
\caption{
Qualitative comparison of specular components between our method and the existing methods. 
}
\label{fig:specularshading_cg}
\end{figure}

Figure~\ref{fig:specularshading_cg} shows a comparison of specular shading with the existing methods. For the method by Pandey \etal~\cite{SIG21_Pandey}, we computed a pseudo-specular output by subtracting a diffuse-only relit image generated using a diffuse light map from a final relit image.
\cite{EGSR21} reproduces gloss using fourth-order spherical harmonics but fails to adequately approximate the high-frequency components of the environmental lighting.
On the other hand, the methods approximating gloss using neural networks~\cite{CVPR20_nestmeyer,PG21,SIG21_Pandey} struggle to learn gloss effectively, leading to blurry output.
This can be also observed in their large standard deviations.
In contrast, our method exhibits high-frequency specular reflection on skin regions in the results.
The quantitative comparison in Table~\ref{tab:specularshading_cg} also shows that our method achieves the best performance in all metrics. 

\begin{table}[t] 
\caption{
Quantitative comparison of specular shading (mean$\pm$standard deviation).
For the method by Pandey \etal~\cite{SIG21_Pandey}, we compute a pseudo-specular output from the difference between a relit image and a diffuse-only relit image. 
} 
\label{tab:specularshading_cg}
\hbox to\hsize{\hfil
\centering
\resizebox{\columnwidth}{!}{
\begin{tabular}{l|ccc}
    \hline
                            & RMSE$\downarrow$                & SSIM$\uparrow$                & LPIPS$\downarrow$ \\
    \hline \hline
    \cite{CVPR20_nestmeyer} & 0.0506\small{$\pm$0.0530}             & 0.741\small{$\pm$0.124}             & \underline{0.0813}\small{$\pm$0.0155}\\ 
    \cite{EGSR21}           & 0.0564\small{$\pm$0.0517}            & 0.460\small{$\pm$0.163}             & 0.208$\pm$\small{0.0428}\\  
    \cite{PG21}             & 0.0484\small{$\pm$0.0345}            & 0.406\small{$\pm$0.172}             & 0.107\small{$\pm$0.0155}\\  
    \cite{SIG21_Pandey}     & 0.0484\small{$\pm$0.0343}            & 0.479\small{$\pm$0.155}             & 0.0910\small{$\pm$0.0170}\\ 
    \hline
    Ours                    & \textbf{0.0334}\small{$\pm$0.0262}    & \textbf{0.835}\small{$\pm$0.0908}   & \textbf{0.0698}\small{$\pm$0.0127}\\
    \hline
    Blinn-Phong             & \underline{0.0390}\small{$\pm$0.0295} & \underline{0.746}\small{$\pm$0.106} & 0.0898\small{$\pm$0.0132}\\ 
    \hline
\end{tabular}\hfil}
}
\end{table}

\subsubsection{Comparison of relighting results}
\label{para:comparison_relit}
\begin{figure*}[t]
\centering
\includegraphics[width=1\linewidth]{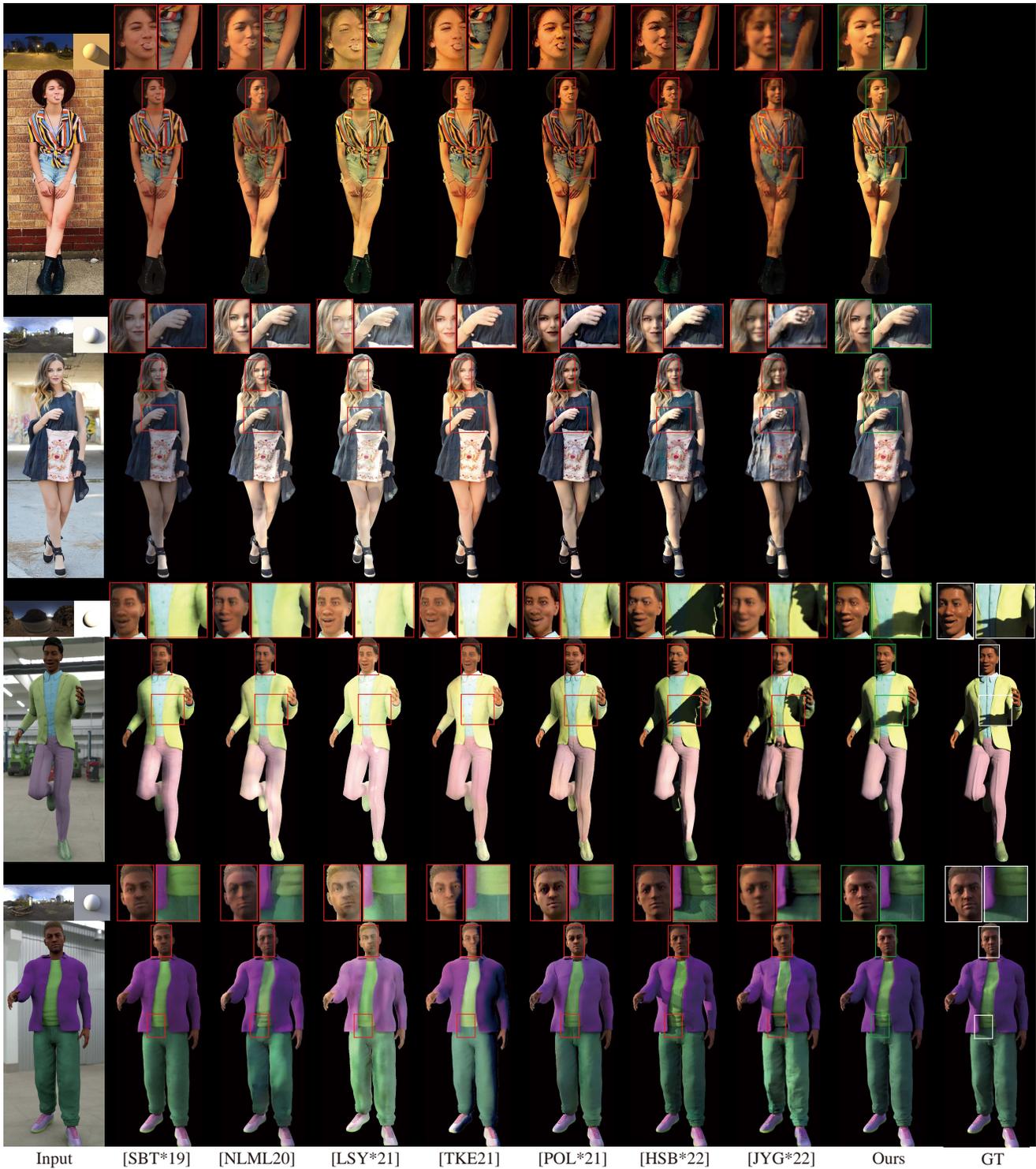}
\caption{
Qualitative comparison of relighting results. The top two rows show the results for real photographs, whereas the bottom two rows show the results for synthetic data. 
Note that there are no ground-truth relit images for the real photographs. 
}
\label{fig:relit_cg_real}
\end{figure*}

\begin{figure*}[t]
\centering
\includegraphics[width=0.948\linewidth]
{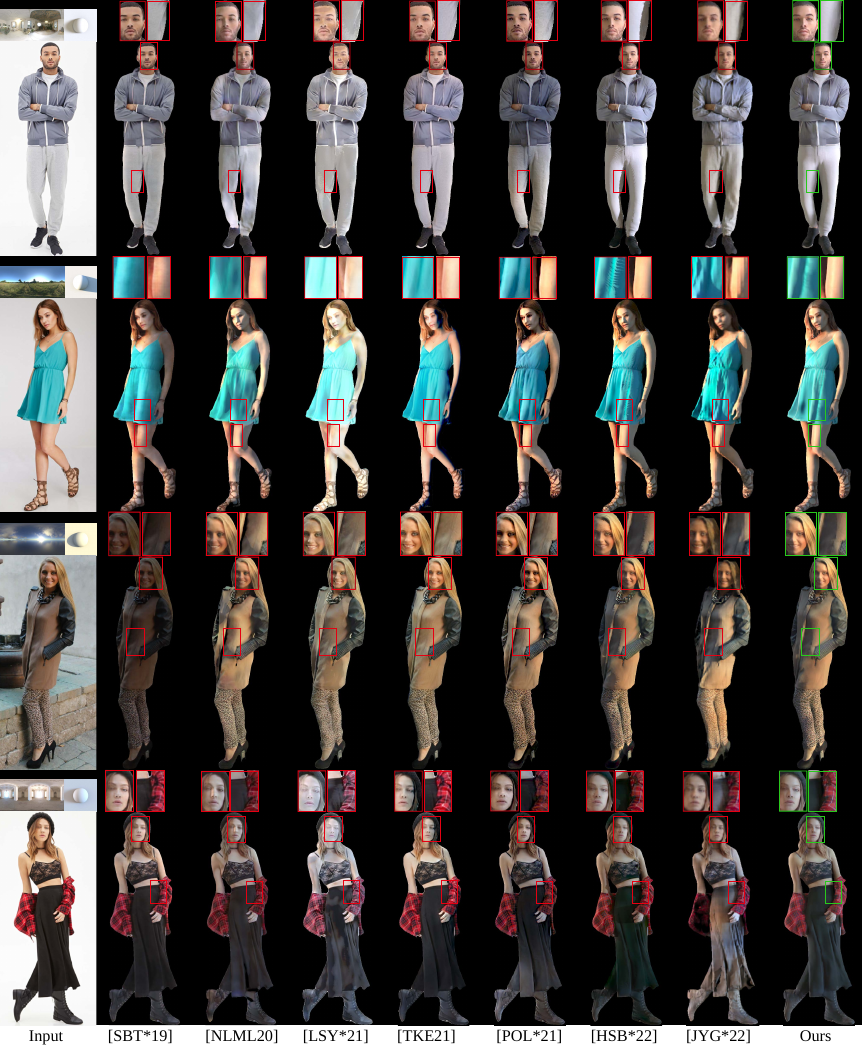}
\caption{
Qualitative comparison of relighting results using the SHHQ dataset~\cite{ECCV22_fu}. 
Note that there are no ground-truth relit images for the real photographs.
}
\label{fig:relit_shhq_real}
\end{figure*}

Table~\ref{tab:relit_cg} and Figures~\ref{fig:relit_cg_real} and \ref{fig:relit_shhq_real} show the quantitative and qualitative comparisons of relighting results between our method and existing methods, respectively. The real images in Figure~\ref{fig:relit_cg_real} were taken from Unsplash, while Figure~\ref{fig:relit_shhq_real} used the SHHQ dataset~\cite{ECCV22_fu}.
For the real photograph inputs, there is no ground-truth relit image. 
The shadows and highlights in the input images do not affect the relighting results so much, thanks to the clean synthetic training data. This benefit is shared not only by our method but also by other methods trained on the same dataset.
The methods~\cite{SIG19,CVPR20_nestmeyer} that do not consider the physical laws overall yielded blurry outputs. 
Because the studies~\cite{EGSR21,PG21} use illumination approximated with low-order SH, they struggle to approximate high-frequency illumination, and the contrast of the shading becomes strong. Because of the same reason, these methods cannot also reproduce high-frequency highlights and shadows. 
The method by Pandey \etal~\cite{SIG21_Pandey} handles low-frequency highlights but fails to reproduce high-frequency highlights and shadows. 
\cite{CVPR22} uses directional light sources to represent illumination and cannot handle low-frequency shadows, resulting in noticeable artifacts around shadow boundaries. In addition, the mesh recovered from the depth map lacks invisible geometry, resulting in inaccurate shadows. In addition, this method is limited to diffuse reflections and cannot reproduce gloss.
\cite{ECCV22} can also reproduce high-frequency shadows by ray-tracing, but artifacts are noticeable. This is due to the presence of many defects and artifacts in the 3D reconstruction results by PIFuHD~\cite{CVPR20_saito}. In addition, it is limited to diffuse reflections and cannot reproduce gloss.
Our method can generate high-frequency shadows due to the light occluded by hands and legs, as well as natural skin highlights, 
by performing physically-based lighting with specular component and 3D geometry. More results are contained in Appendix~\ref{appendix:relit_moreresult}.

To further validate our method, we conducted a user study with 20 real images collected from Unsplash and lit under natural illumination. We compared four methods with the top scores with synthetic data. 22 subjects were requested to pick the most natural relighting result for each input image, emphasizing facial highlights and shadows around arms and clothes. 
Consequently, the selection percentages are: \cite{SIG19} 4.1\%, \cite{SIG21_Pandey} 16.6\%, \cite{CVPR22} 18.6\%,
and ours 60.7\%, which means that ours outperforms other methods. 
The relighting results used in the user study are published in Appendix~\ref{appendix:relit_userstudy}.

\begin{table}[t] 
\caption{Quantitative comparison of relighting results (mean$\pm$standard deviation).} 
\label{tab:relit_cg}
\hbox to\hsize{\hfil
\centering
\resizebox{\columnwidth}{!}{
\begin{tabular}{l|ccc}
    \hline
    & RMSE$\downarrow$ & SSIM$\uparrow$ & LPIPS$\downarrow$\\
    \hline \hline
    \cite{SIG19} & 0.122\small{$\pm$0.0944} & 0.712\small{$\pm$0.148} & 0.0556\small{$\pm$0.0196}\\ 
    \cite{CVPR20_nestmeyer} & 0.137\small{$\pm$0.0843} & 0.678\small{$\pm$0.117} & 0.0717\small{$\pm$0.0157}\\ 
    \cite{EGSR21} & 0.224\small{$\pm$0.122} & 0.483\small{$\pm$0.149} & 0.162\small{$\pm$0.0437}\\  
    \cite{PG21} & 0.173\small{$\pm$0.0980} & 0.586\small{$\pm$0.166} & 0.0737\small{$\pm$0.0206}\\  
    \cite{SIG21_Pandey} & 0.111\small{$\pm$0.0645} & 0.692\small{$\pm$0.116} & 0.0538\small{$\pm$0.0145}\\
    \cite{CVPR22} & 0.101\small{$\pm$0.0723} & 0.632\small{$\pm$0.120} & 0.0598\small{$\pm$0.0235}\\ 
    \cite{ECCV22} & 0.121\small{$\pm$0.0650} & 0.686\small{$\pm$0.110} & 0.0819\small{$\pm$0.0132}\\ 
    \hline
    Ours (full) & \textbf{0.0744}\small{$\pm$0.0411} & \textbf{0.787}\small{$\pm$0.0973} & \textbf{0.0493}\small{$\pm$0.0131}\\  
    \hline
    Ours w/o specular & \underline{0.0780}\small{$\pm$0.0415} & \underline{0.771}\small{$\pm$0.0972} & \underline{0.0502}\small{$\pm$0.0133}\\ 
    Ours w/o refinement or specular& 0.0882\small{$\pm$0.0479} & 0.757\small{$\pm$0.100} & 0.0528\small{$\pm$0.0135}\\  
    Ours w/o shadow or specular& 0.0946\small{$\pm$0.0559} & 0.753\small{$\pm$0.103} & 0.0518\small{$\pm$0.0138}\\
    \hline
\end{tabular}\hfil}
}
\end{table}

\subsubsection{Evaluation of temporal consistency}
We evaluated the temporal consistency for relit videos obtained using dynamic lights. For 10 static 3DCG human data of different identities, we generated 20 videos consisting of 128 frames by rotating two randomly-selected test environment maps in the longitude direction. 

A typical evaluation metric for temporal consistency uses a warping function based on optical flow. 
However, the warping function is inappropriate for static humans relit with dynamic lights.
Regarding evaluation metrics,
\cite{DBLP:journals/cvm/LiuZCWLQD24} proposed the color distribution consistency index (CDC) as a measure of temporal consistency when applying a colorization task to each video frame.
CDC calculates the similarity of the color distribution between consecutive frames.
However, CDC cannot handle dynamically changing color distributions due to dynamic lighting and prefers blurry frames and frames without shading changes because CDC does not refer to ground truth.
Therefore, inspired by CDC, we use the following metric to evaluate frame differences within different frame intervals with reference to ground truth:

\begin{align}
\label{eq:unique_flicker_metrics_1}
\partial \mathrm{RMSE}_t &= \frac{1}{N_f-t} \mspace{-7mu} \displaystyle \sum_{i=1}^{N_f-t}\mathrm{RMSE}(\mathbf{F}_{i+t}-\mathbf{F}_{i},\hat{\mathbf{F}}_{i+t}-\hat{\mathbf{F}}_{i}), \\[-0.2em]
\label{eq:unique_flicker_metrics_2}
\partial \mathrm{RMSE} &= \frac{1}{3}(\partial\mathrm{RMSE}_1+\partial\mathrm{RMSE}_2+\partial\mathrm{RMSE}_4),
\end{align}
where $\mathbf{F}$, $N_{f}$, and $t$ are the ground-truth relighting frames, the total number of frames, and the time step, respectively.
As with CDC, the use of different time steps allows for long- and short-term temporal consistency.
The second column of Table~\ref{tab:flicker_unique} shows the results.
We can see that our relighting results faithfully reproduce the changes in shading, yielding a temporally consistent output.

Furthermore, we quantitatively evaluated the temporal consistency of relighting results with synthetic dynamic people by preparing eight animation sequences with 
four rigged human models animated using Mixamo~\cite{mixamo}
and a pair of environment maps for before and after relighting.
The third column of Table~\ref{tab:flicker_unique} shows the results.
Ours yields the least flickering with the best fidelity to the ground truth.

\begin{table}[t] 
\caption{
Quantitative comparison of temporal consistency for dynamic lights and people using our metric (Equation~\eqref{eq:unique_flicker_metrics_2}). 
} 
\label{tab:flicker_unique}
\hbox to \hsize{\hfil
\centering
\small
\begin{tabular}{l|cc}
    \hline
    & \multicolumn{2}{c}{$\partial \mathrm{RMSE} \downarrow$} \\ \cline{2-3}
    & Dynamic lights & Dynamic people \\
    \hline \hline
    \cite{SIG19} & 0.473 & 0.140 \\
    \cite{CVPR20_nestmeyer} & 0.447 & 0.123\\
    \cite{EGSR21} & 0.530 & 0.176\\
    \cite{PG21} & 0.470 & 0.132\\
    \cite{SIG21_Pandey} & 0.394 & \underline{0.116}\\
    \cite{CVPR22} & 0.399 & 0.121\\
    \cite{ECCV22} & \underline{0.388} & \underline{0.116}\\ 
    Ours & \textbf{0.386} & \textbf{0.113}\\   
    \hline
\end{tabular}\hfil}
\end{table}

\section{Conclusions}
\label{sec:conclusions}
We have proposed a relighting method for full-body images of clothed humans, taking into account low- to high-frequency specular reflections and shadows.
Our method first applies inverse rendering to the input human image to obtain diffuse and specular reflectance maps, roughness map, and depth map, and then reproduces both low- and high-frequency gloss and shadows based on the Disney principled BRDF and convolutional shadow mapping. 
A new lighting representation with a fixed number of area lights was proposed to implement it.
The experimental results revealed that our method reproduces shadows and highlights more plausibly than existing methods that approximate shadows and highlights using neural networks, demonstrating the effectiveness of formulating highlights and shadows based on physically-based lighting.

Currently, our method does not explicitly model complicated lighting effects such as anisotropic reflection and subsurface scattering.
In future work, we would like to approximate them using physics-aware neural networks or approaches to bridging synthetic-to-real domain gap~\cite{PG21,SIGA22}.


{
    \small
    \bibliographystyle{eg-alpha-doi}
    \bibliography{main}
}

\newpage
\appendix

\section{Candidate Architectures for Depth Estimation Network}
\label{appendix:DepthEstimationNetwork}

We attempted the following three network architectures for estimating the depth map from a single full-body human image in this study.

\renewcommand{\labelenumi}{(\alph{enumi})}
\begin{enumerate}
    \item PIFu~\cite{ICCV19_saito}-based network
    \item U-Net~\cite{MICCAI15}
    \item Network with kernel size used for convolution of U-net to 1 only for the last output layer.
\end{enumerate}
\renewcommand{\labelenumi}{(\arabic{enumi})}

Among the options above, (a) represents a network structure based on PIFu, a method for reconstructing a 3D model of a person from a single full-body image of the person. This method achieves 3D shape reconstruction by estimating the implicit function that represents the 3D person. Building upon the success of this 3D shape reconstruction, the network structure is modified to estimate the depth information of the person instead of the implicit function and trained from scratch using our dataset. (b) employs U-Net~\cite{MICCAI15}, a network commonly used in image-to-image translation. 
(c) is inspired by PIFu's pixel-aligned concept
to faithfully reflect image features and is designed to strongly emphasize per-pixel features by adjusting the kernel size used in the convolution of the final output layer in (b) to one.
As a preliminary experiment, quantitative comparisons are presented in Table~\ref{tab:net_2nd_net}. The preliminary experiments demonstrate that (c) performs the best, and therefore, (c) is adopted in the subsequent experiments.

\begin{table}[t] 
\caption{
Quantitative comparison of depth map estimation with the three network architectures. 
} 
\label{tab:net_2nd_net}
\hbox to\hsize{\hfil
\centering
\small
\begin{tabular}{l|ccc}
    \hline
    & RMSE$\downarrow$ & SSIM$\uparrow$ & LPIPS$\downarrow$\\
    \hline \hline
    Network(a) & 0.0351 & 0.991 & 0.00361\\
    Network(b) & \underline{0.0209} & \underline{0.994} & \underline{0.00264}\\
    Network(c) & \textbf{0.0202} & \textbf{0.995} & \textbf{0.00260}\\
    \hline
\end{tabular}\hfil}
\end{table}

\section{Comparison with Different Numbers of Area Lights}
\label{appendix:NoOfLights}
\begin{figure}[t]
\centering
\includegraphics[width=1\linewidth]{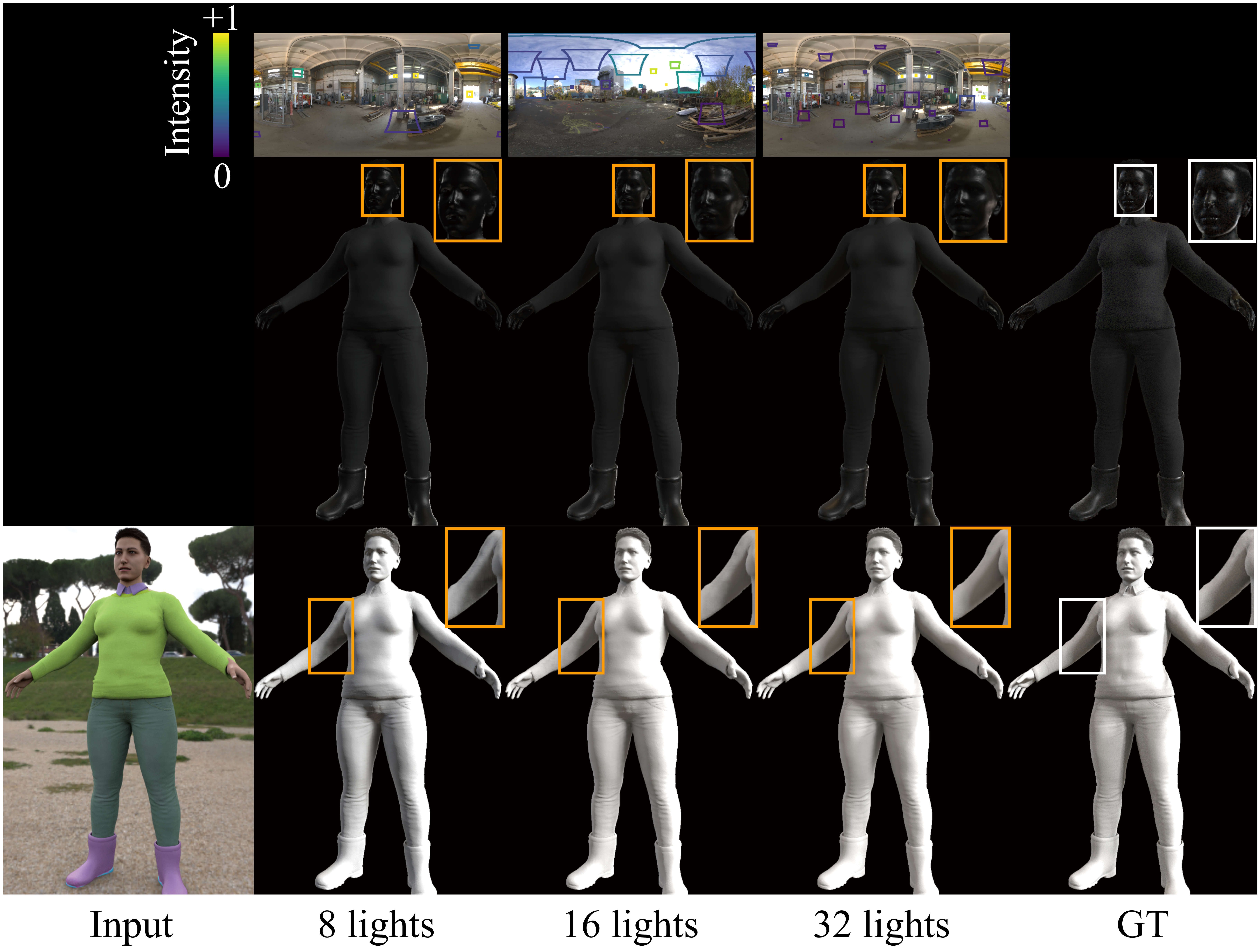}
\caption{
Comparison of shadings and shadows for different numbers of light sources.
The top row shows a visualization of optimized light parameters on an environment map, the middle row shows specular shading with shadow, and the bottom row shows diffuse shading with shadow.
}
\label{fig:n_light}
\end{figure}

We conducted evaluations with different numbers of area lights, $N_L = 8, 16, 32$.
A qualitative comparison is shown in Figure~\ref{fig:n_light}, and a quantitative comparison of the relighting results and inference times is shown in Table~\ref{tab:n_light}.
In the quantitative comparison, the accuracy is higher when 32 lights are used for all metrics except LPIPS, but the inference time is about three times longer than when 16 lights are used, which cannot be ignored for training and testing.
In the qualitative evaluation, some artifacts were observed when 8 lights were used, but by passing through the shadow refinement network, shadows close to the ground truth were reproduced even with a small number of light sources.
There was little change in highlights between 16 and 32 lights.

\begin{table}[t] 
\caption{Quantitative comparison of relighting results and inference times with different numbers of area lights. 
}
\label{tab:n_light}
\hbox to\hsize{\hfil
\centering
\resizebox{\columnwidth}{!}{
\begin{tabular}{l|cccc}
    \hline
    & RMSE$\downarrow$ & SSIM$\uparrow$ & LPIPS$\downarrow$  & \makecell{Inference\\time (sec.)}\\
    \hline \hline
    8 lights         & 0.0903\small{$\pm$0.0527}          & 0.761\small{$\pm$0.100}           & 0.0517\small{$\pm$0.0146} & \textbf{0.682}\\
    16 lights (Ours) & \underline{0.0744}\small{$\pm$0.0411}          & \underline{0.787}\small{$\pm$0.0973}          & \textbf{0.0493}\small{$\pm$0.0131} & \underline{1.01}\\
    32 lights        & \textbf{0.0729}\small{$\pm$0.0409} & \textbf{0.790}\small{$\pm$0.0972} & \underline{0.0500}\small{$\pm$0.0132} & 3.20\\
    \hline
\end{tabular}\hfil}
}
\end{table}

\section{Qualitative results of user study}
\label{appendix:relit_userstudy}
\begin{figure*}[t]
\centering
\includegraphics[width=1\linewidth]{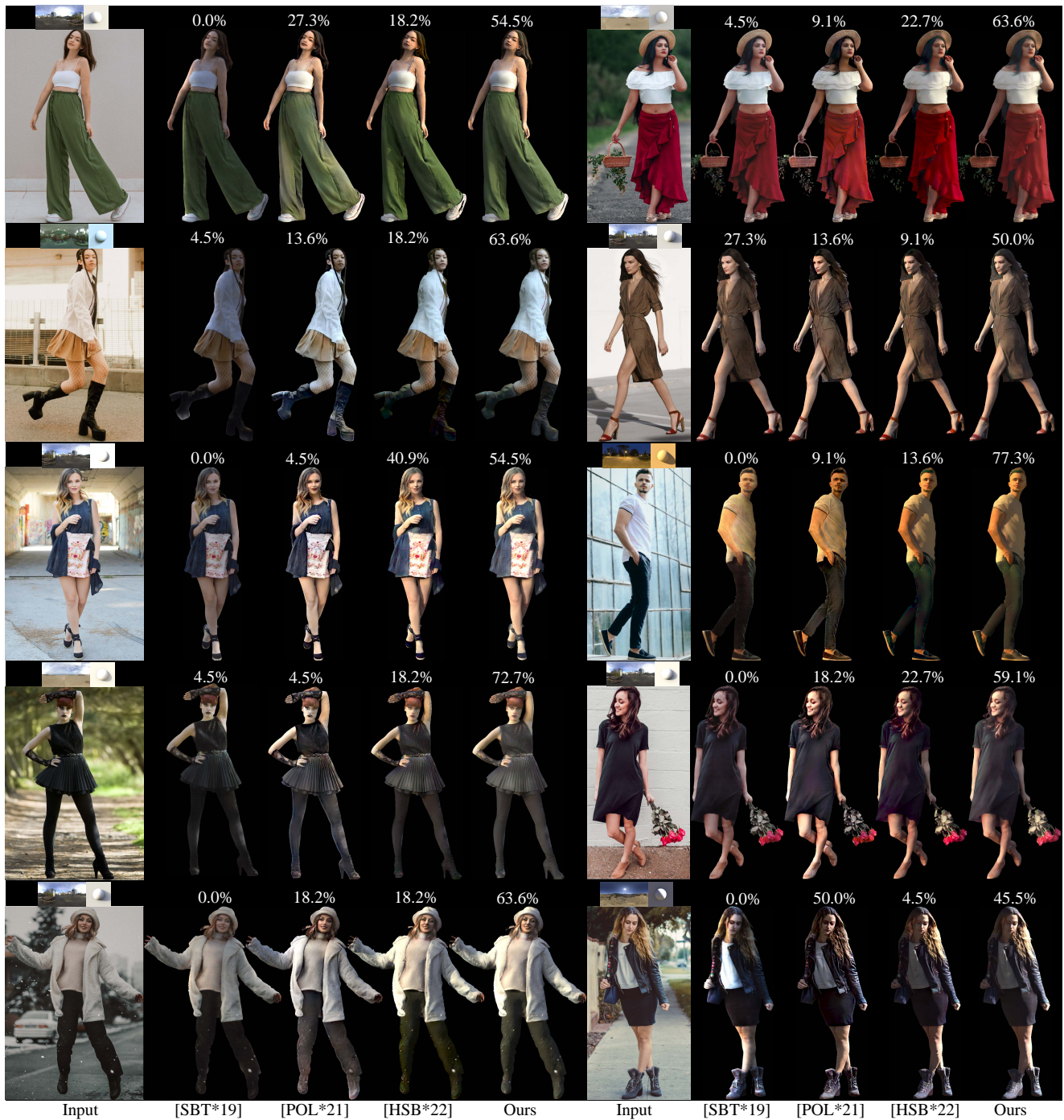}
\caption{
Relighting results that were shown to the subjects in the user study.
The percentage above each result indicates the portion of the 22 subjects who selected the image as the most photorealistic.
}
\label{fig:relit_userstudy_1}
\end{figure*}

\begin{figure*}[t]
\centering
\includegraphics[width=1\linewidth]{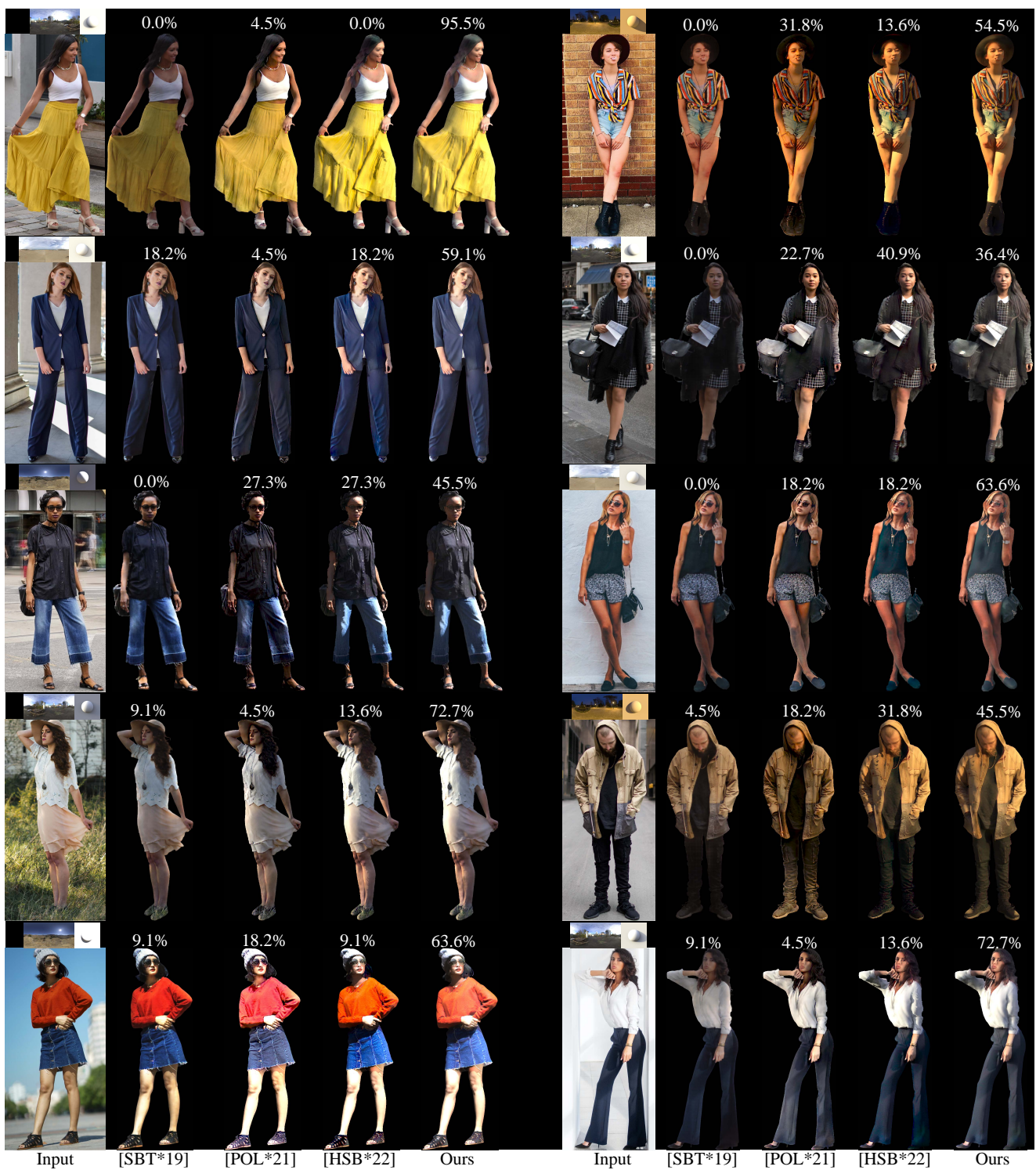}
\caption{
Relighting results that were shown to the subjects in the user study.
The percentage above each result indicates the portion of the 22 subjects who selected the image as the most photorealistic.
}
\label{fig:relit_userstudy_2}
\end{figure*}

Figures~\ref{fig:relit_userstudy_1} and \ref{fig:relit_userstudy_2} show the relighting results of the real images used in the user study. We used 20 real images collected from Unsplash. We compared the four methods with the highest scores in the quantitative evaluation using synthetic data.

\section{Additional results}
\label{appendix:relit_moreresult}
\begin{figure*}[t]
\centering
\includegraphics[width=0.813\linewidth]{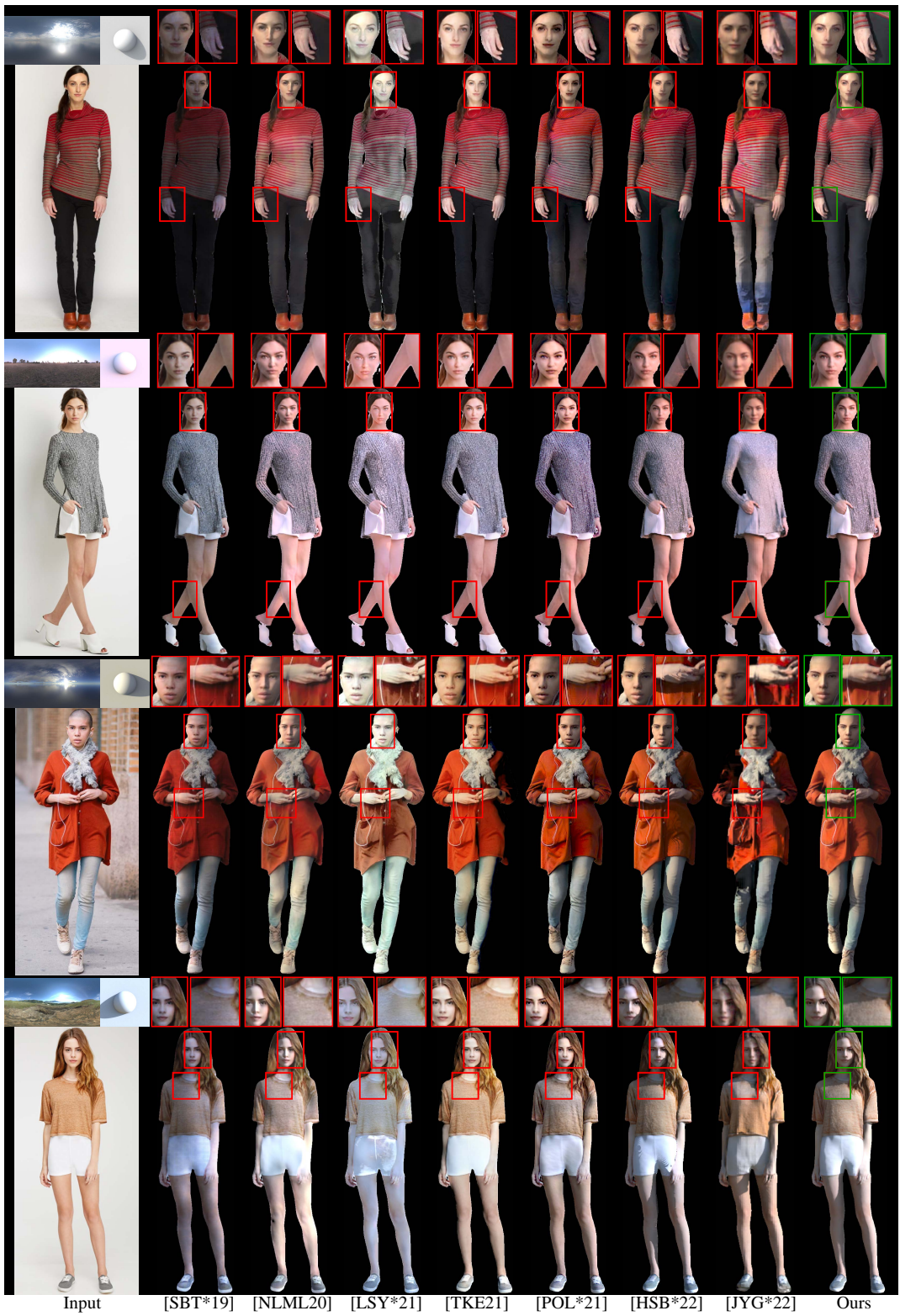}
\caption{
Qualitative comparison of relighting results using the SHHQ dataset~\cite{ECCV22_fu}.}
\label{fig:relit_shhq_1}
\end{figure*}

\begin{figure*}[t]
\centering
\includegraphics[width=0.806\linewidth]{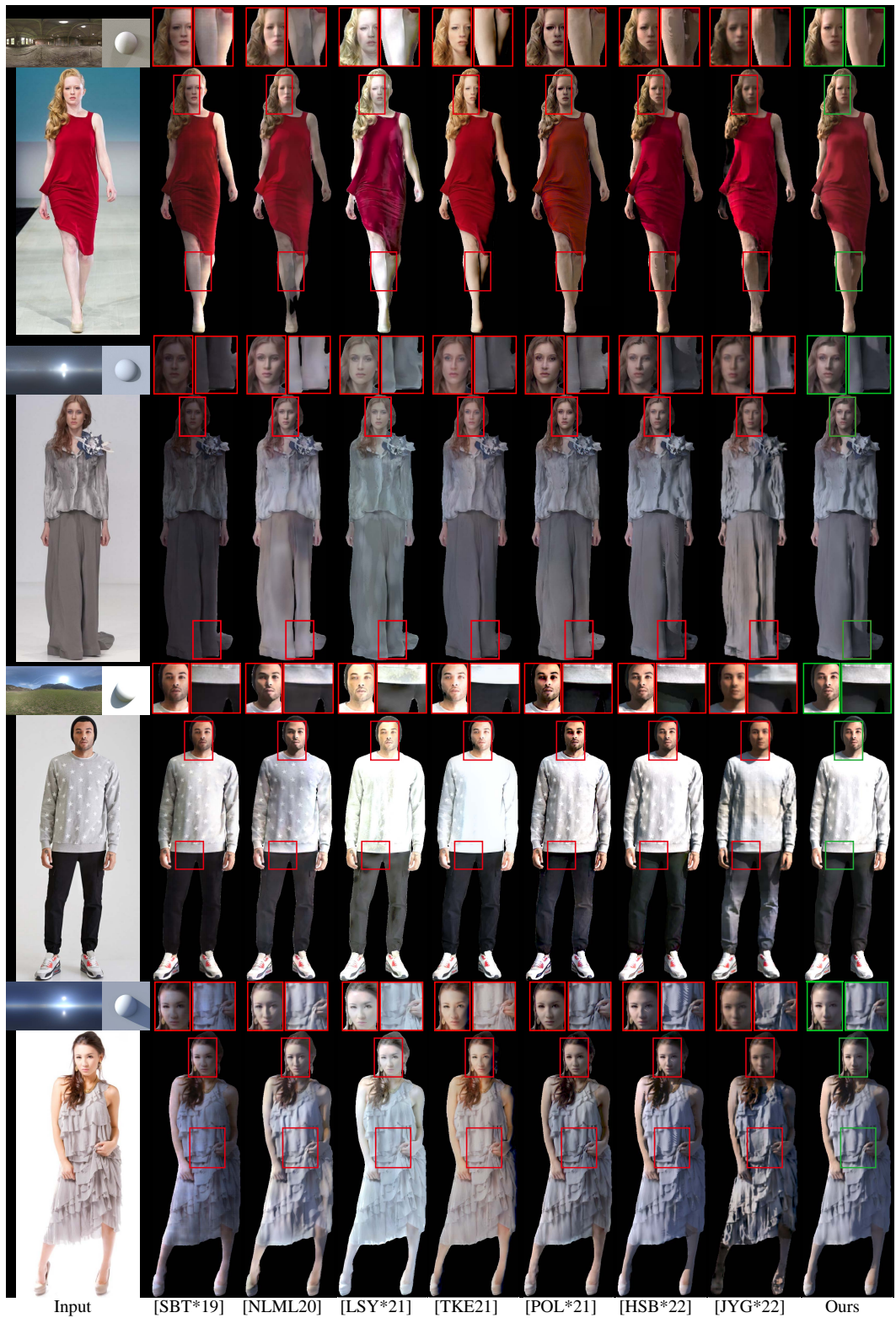}
\caption{
Qualitative comparison of relighting results using the SHHQ dataset~\cite{ECCV22_fu}.}
\label{fig:relit_shhq_2}
\end{figure*}

Figures~\ref{fig:relit_shhq_1} and \ref{fig:relit_shhq_2} show more relighting results with the SHHQ dataset~\cite{ECCV22_fu}. 


\end{document}